\documentclass[11pt]{article}

\usepackage{graphicx,amsmath}
\usepackage[authoryear]{natbib}
\usepackage[font={small},bf]{caption}
\usepackage[letterpaper,margin=1in]{geometry}
\usepackage{float}
\usepackage{array}

\graphicspath{ {figures/} }

\newcommand{\ud}{\text{d}}

\let\originalleft\left
\let\originalright\right
\renewcommand{\left}{\mathopen{}\mathclose\bgroup\originalleft}
\renewcommand{\right}{\aftergroup\egroup\originalright}

\begin{document}

\title{Translucent windows: how uncertainty in competitive interactions impacts detection of community pattern}
\author{Rafael D'Andrea$^{1, \ast}$, Annette Ostling$^{2, \ast\ast}$ \& James O'Dwyer$^{1, \dagger}$ \\ 
  $^1${\small Dept.~Plant Biology, University of Illinois at Urbana-Champaign, IL, USA}\\
  $^2${\small Dept.~Ecology \& Evolutionary Biology, University of Michigan, Ann Arbor, MI, USA}\\  $^{\ast}${\small \texttt{rdandrea@illinois.edu}}, \small{\texttt{ORCID 0000-0001-9687-6178}}\\ $^{\ast\ast}${\small \texttt{aostling@umich.edu}}, $^{\dagger}${\small \texttt{jodwyer@illinois.edu}} 
}
\date{}

\maketitle


\begin{abstract}
Trait variation and similarity among coexisting species can provide a window into the mechanisms that maintain their coexistence. Recent theoretical explorations suggest that competitive interactions will lead to groups, or clusters, of species with similar traits. However, theoretical predictions typically assume complete knowledge of the map between competition and measured traits. These assumptions limit the plausible application of these patterns for inferring competitive interactions in nature. Here we relax these restrictions and find that the clustering pattern is robust to contributions of unknown or unobserved niche axes. However, it may not be visible unless measured traits are close proxies for niche strategies. We conclude that patterns along single niche axes may reveal properties of interspecific competition in nature, but detecting these patterns requires natural history expertise firmly tying traits to niches.
\end{abstract}


\textbf{Corresponding author}: Rafael D'Andrea, rdandrea@illinois.edu, (631) 645-5710, 172 Morrill Hall 505 S Goodwin Ave, Urbana, IL, 61801 \newline

\textbf{Data accessibility statement}: This study made use of simulations. The code to generate and analyze the data is available on GitHub, https://github.com/rafaeldandrea/Translucent-windows-2018-code. \newline

\textbf{Statement of Authorship}: RD, AO, and JD conceived the study. RD performed modeling work and analyzed output data. RD wrote the first draft of the manuscript, and all authors contributed substantially to revisions.\newline

\textbf{This is the peer reviewed version of the following article: ``Translucent windows: how uncertainty in competitive interactions impacts detection of community pattern'', which has been accepted for publication at Ecology Letters,  DOI: 10.1111/ele.12946. This article may be used for non-commercial purposes in accordance with Wiley Terms and Conditions for Self-Archiving.}

\newpage
\section*{Introduction}
How does competition shape the distribution of species traits in nature? The classic answer, based on the idea that similarity breeds competition, is that coexisting species will be more different than expected by chance \citep{MacArthur1967, Abrams1983}. Although intuitive, this prediction of limiting similarity lacks widespread empirical support \citep{Chase2003, Gotzenberger2012} despite extensive research \citep{DAndrea2016b}. Recent studies have found that trait-mediated competition can actually lead to the emergence of clusters, or groups of similar species \citep{Bonsall2004, Scheffer2006, Hernandez-Garcia2009, Sakavara2018}. Although contradictory on the surface, clusters are a natural extension of limiting similarity. If left alone, competitive sorting culminates in maximally differentiated species, each occupying a niche. But an influx of new individuals and species due to speciation or immigration may keep the number of extant species above the number of local niches \citep{DAndrea2018}. In that case, species whose traits place them near a niche will persist longer \citep{Scranton2016, DAndrea2017d}, and species in between are quickly excluded as they are not well adapted to any niche \citep{Barabas2013a, Vergnon2013}. The incorporation of species clusters to traditional trait-pattern theory may help explain the mixed evidence for limiting similarity, and may help understand coexistence in species-rich systems such as tropical forests. 

Nevertheless, clusters are not a widely observed phenomenon in nature outside specific communities \citep{Scheffer2006, Scheffer2015, Segura2011, Segura2013, Vergnon2009, Yan2012}. While this can in part be due to clusters having only recently been connected with competition in the theoretical literature \citep{DAndrea2016b}, it is not clear that clusters should occur outside idealized models. Theoretical studies typically make two critical assumptions that do not hold in real communities, thus limiting their applicability. First, they assume that species can be arranged on a line such that distances between their relative positions on the line determine how strongly they compete with one another \citep{MacArthur1967}. In reality, it is unlikely that positions along a single line can fully predict the degree of competition. For example, birds that eat very distinct foods may still compete strongly for nesting sites, or may be under apparent competition because they share predators or parasites. Knowledge of the birds' diets will predict degree of competition better than chance, but the relationship will appear noisy because of the contribution of these other factors. In this study we will refer to the line as a niche axis, such that separation on that axis predicts the degree of competition between species. From the vantage point of the food-niche axis, the unknown contributions of the nesting site axis, predator axis, etc loosen the link between diet and competition. While there is theoretical evidence that multidimensional niche space may cause multidimensional niche clustering \citep{Fort2010}, it is unclear whether any pattern should be expected along a single niche axis in the presence of these other unobserved factors.

The second assumption is that measurable traits such as body size and leaf tissue density are actual niche axes, so patterns driven by competition should be visible as patterns in the distribution of these traits. However, there is a conceptual distinction between aspects of phenotype directly responsible for competitive interactions and aspects of phenotype that are typically measured. The former may form niche axes, but are unlikely to perfectly coincide with easily measurable morphological or physiological traits (Appendix A). At best, measured functional traits are good predictors of the true causes of competitive interactions \citep{McGill2006, Violle2007}, but that map is inevitably noisy. For example, consider plant strategies regarding light capture and shade tolerance, or the ``light-niche'' axis. Leaf tissue density correlates with shade tolerance in forests, but will not fully predict the light-niche because the latter is also affected by many other plant traits, such as maximum plant height and wood density \citep{Wright2010}. Pattern in the distribution of light strategies across species will translate into pattern in the distribution of leaf tissue density only to the extent that the latter reflects the former. 

\begin{figure}[H]
\caption{\textbf{Causal links between traits and community structure}. In nature (top), the full set of functional traits completely determines strategies in niche space; the latter in turn consists of multiple niche axes, the full set of which completely determines the competitive interactions between species; competition drives clustering in niche space, which translates as measurable clustering in trait space. A complete model of nature would require impossibly complete knowledge and measurement of all the relevant elements and links. An effective model (bottom) focuses on a subset of trait and niche axes, and treats the unmeasured/unmodeled contribution of all other factors as noise. As an example, we show an effective model where a single niche axis (light-related strategy) is an imperfect predictor of competitive interactions, with the unknown contribution of all other niche axes comprising process noise; in turn, a single functional trait (maximum height) is an imperfect predictor of the light-niche strategy, and the unknown contribution of all other functional traits enters as measurement noise.
}\label{Fig1}
\includegraphics[width=1\textwidth,angle=0]{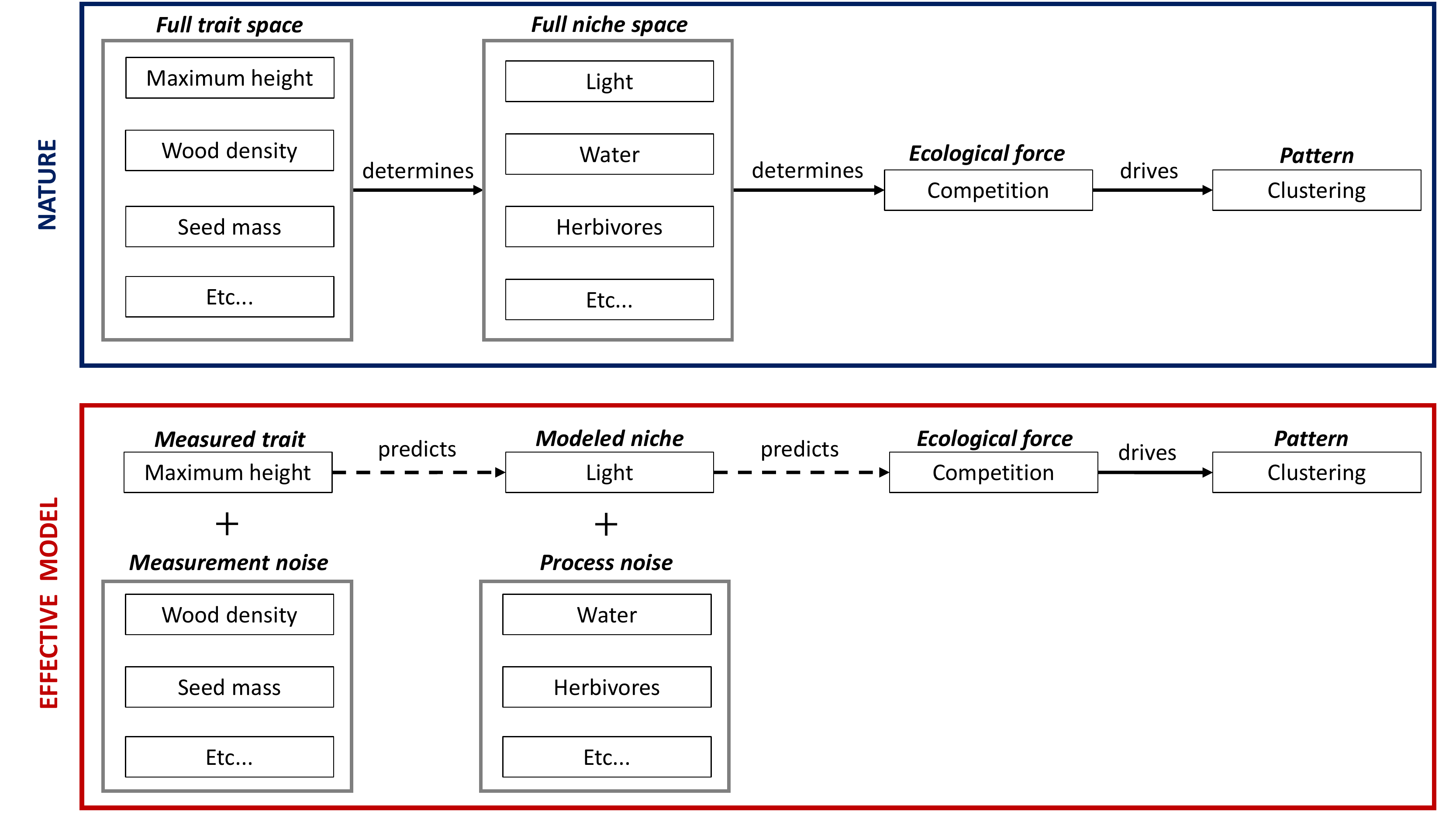}
\centering
\end{figure}

Little is known about how this inevitable uncertainty might affect our ability to detect pattern in nature (Fig.~\ref{Fig1}). If the clusters predicted by theory are robust to uncertainty, then they are useful for diagnosing competition based on overlap in niche strategies. But if measuring an incomplete or inaccurate set of traits means we are unlikely to detect clusters, then there is little value in searching for them in nature. 

Here we test this robustness. We start by asking what happens when only one of the niche axes responsible for competitive interactions is fully known. We find that pattern on one niche axis is surprisingly robust to the contributions of other niche axes. Next we ask whether this niche pattern translates well into trait pattern, given that traits are imperfect predictors of niches. We find that the pattern is quite sensitive to this error, and its detectability quickly fades as we loosen the link between niche axis and measured trait. We then show that measuring multiple traits can provide more accurate estimates and thus reduce measurement noise, but ultimately a tight functional relation between traits and niche strategies is key.

\section*{Materials and Methods}

\subsection*{Niche dynamics}
Many ecological communities consist of multiple species competing for resources, and often these resources are thought of as falling on a continuous gradient. One such example is seed-eating birds that may feed on seeds of various sizes. The species niche reflects its resource preferences within the gradient, so that similarity in niche values denotes the amount of overlap between those strategies \citep{MacArthur1967, Leimar2013}. 

We represent this scenario using a Lotka-Volterra model of competitive dynamics. 
\begin{equation}
\frac{\ud N_i}{\ud t}=r_i\left(1-\sum_j A_{ij}\frac{N_j}{K_i}\right)N_i+m_i\nonumber
\end{equation}
where $N_i, r_i, K_i$ are the abundance, intrinsic (maximal) growth rate, and carrying capacity of species $i$, and the competition coefficients $A_{ij}$ quantify the per-capita competitive impact of species $j$ on species $i$. Steady immigration $m_i$ maintains all abundances above zero. 

We use $S=200$ species in our simulations, and let species differ only by niche strategy, affecting $A_{ij}$. We set parameters $r=1$, $K=100$, and $m=0.2$ immigrating individuals per species per year, identical across all species. These parameter choices lead to equilibrium community sizes between 600 and 35,000 individuals. As a reference point, the 50-hectare forest plot in Barro Colorado Island has approximately 300 species, 20,000 trees, and an estimated $m=0.1$ immigration event per birth \citep{Hubbell2001}. We verified that our results are robust to a four-fold change in immigration rates. All simulations are performed using library deSolve in the R language \citep{RCoreTeam2017}.

\subsection*{Competition coefficients---the kernel}

We will refer to the matrix of competition coefficients $A_{ij}$ as the competition kernel, by analogy with continuous formulations. We set
\[
A_{ij} = a_0\exp(-d_{ij}^4)
\]
where $d_{ij}$ is the distance in niche space between species $i$ and $j$, and the scaling factor $a_0$ is some constant. This quartic function is a generic monotonic shape commonly used in competition models \citep{Pigolotti2010, Leimar2013}. A plot of this function reveals two distinct zones: a plateau of intense competition between similar species, and a steep drop-off beyond a threshold niche distance (Fig.~S3A). We refer to the inner region as the ``core'' of the kernel, and the outer region as its ``tail''. As we will show, this core-tail structure is instrumental in pattern-formation. The threshold distance is somewhat arbitrary, and here we take it to be the niche distance where the kernel is equal to its mean value. Other choices do not qualitatively change our results. 

\subsection*{Process noise: competition in two-dimensional niche space}
\begin{figure}[H]
\caption{\textbf{Hidden niche dimensions add process noise; imperfect traits add measurement noise}. \emph{Process noise} \textbf{A}: Hypothetical niche space formed by two complementary niche axes, with species labeled in order of Euclidean xy-distance to the black species. \textbf{B}: Intensity of competition with black species has a monotonic relationship with the two-dimensional distance. \textbf{C}: Positions of the same species along the x-niche axis, with species labeled in order of x-distance to black. \textbf{D}: Plot of competitive interactions with black as a function of x-distance shows a non-monotonic relationship. The mismatch between B and D is process noise, which is commensurate with the relative importance of the unknown contributions of the y-niche. \emph{Measurement noise} \textbf{E}: Measured traits estimate unseen niche strategies, but the map is imperfect, and relative positions may switch between niche and trait axes. \textbf{F}: Arranging species in increasing order of niche-distance to species 1 gives a monotonic relationship with competition. \textbf{G}: Arranging species in increasing order of trait-distance to species 1 yields an overall decreasing but noisy relationship. The mismatch between F and G is measurement noise, which is commensurate with the imperfections in the niche-trait map.}
\label{Fig2}
\includegraphics[width=.75\textwidth,angle=0]{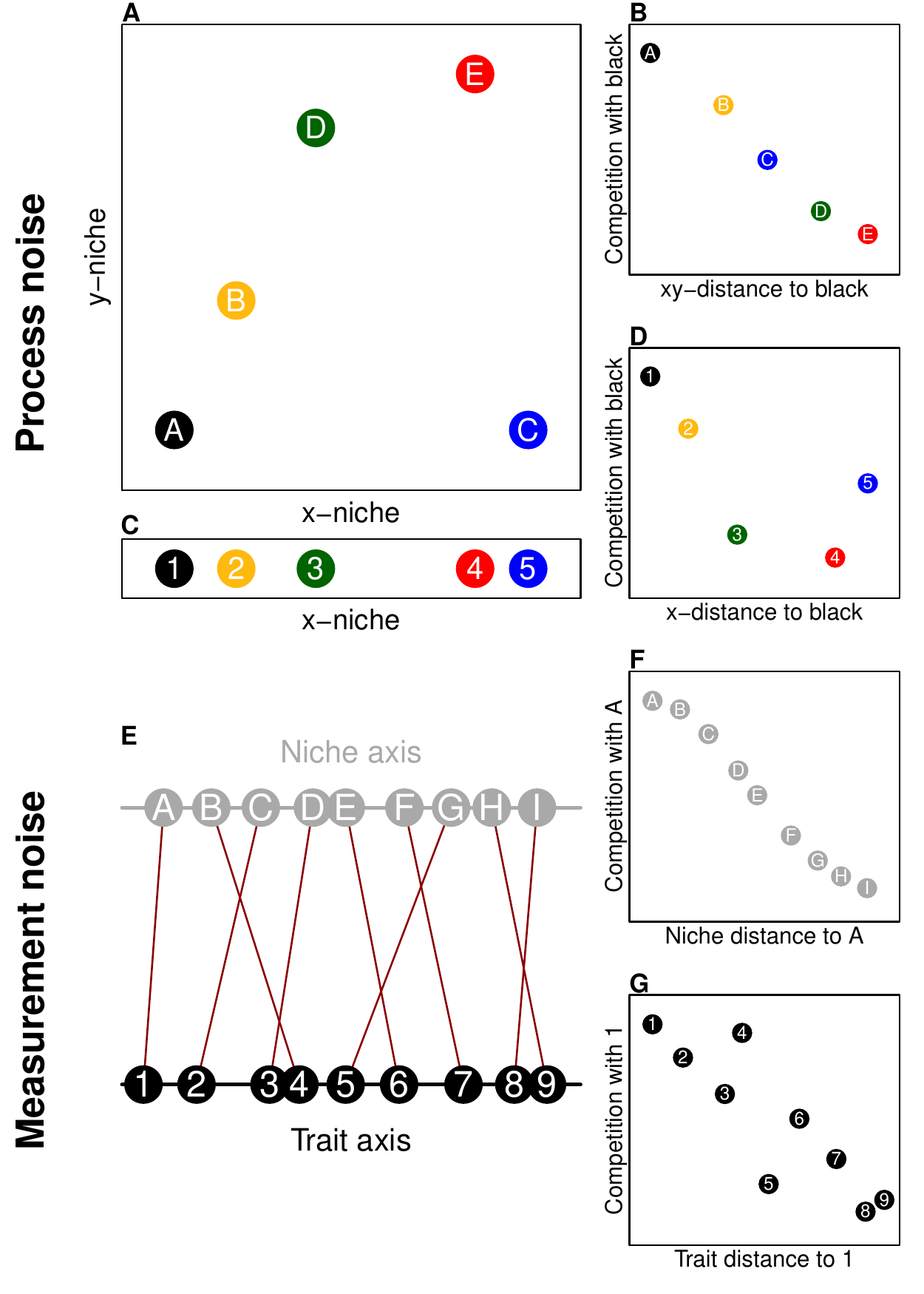}
\centering
\end{figure}

Consider a case where competitive interactions are collectively determined by species strategies along two niche axes, e.g. two different types of resources such as food and nesting sites, or two different types of food such as seeds and insects. Suppose also we only have access to one such axis. The contributions of the unknown niche axis add randomness (i.e. noise) to the otherwise monotonic relationship between competition kernel and niche differences on the known axis (Fig.~\ref{Fig2}). We call this \emph{process noise}.

Two different niche axes can interact in various ways to determine niche overlap \citep{Holt1987}. For example, the two niche axes may reflect strategies for acquiring complementary resources. In this case, similarity on one niche axis (e.g. similar preferences in insect prey) can be compensated via differences along the other axis (foraging for very different seeds). Alternatively, the niche axes may represent fundamentally different needs such as food and nesting sites for birds, or water and light for plants. In this case, no amount of niche differences on one axis can compensate for similarity on the other. We term these scenarios complementary and essential niche axes, respectively. 

In our model, these different scenarios define the two-dimensional niche separation between species. In the case of complementary niche axes, we set 
\[
d_{ij}=\sqrt{\left(\frac{x_i-x_j}{w_x}\right)^2+\left(\frac{y_i-y_j}{w_y}\right)^2}
\]
where $x_i$ ($x_j$) is the x-niche strategy of species $i$ ($j$), which can range from 0 to 1, $w_x$ is a scaling factor that weighs the x-niche difference relative to the y-niche difference, and vice-versa for y-based parameters. Notice that if niche differences on either axis are large, $d_{ij}$ will be large, and therefore the competition coefficient will be small, regardless of differences on the other. 

In the essential niche axes case, we set
\[
d_{ij}=\min\left(\left|\frac{x_i-x_j}{w_x}\right|,\;\left|\frac{y_i-y_j}{w_y}\right|\right)
\]
In this case, $d_{ij}$ will be small if either the x- or y-niche separation is small, regardless of how large the other niche separation may be. Here species cannot compensate for similarity on one axis by dissimilarity on another, and the limiting resource is the one for which they compete the most. 

The constants $w_x$ and $w_y$ define the relative weight of each niche axis to the overall niche separation. In the complementary niche axes scenario, the y-contribution vanishes when $w_y$ is very large relative to $w_x$, and dominates the coefficient when it is very small. In the essential niche axes scenario, the reverse is true. If we can measure the x-niche strategies of our species but have no access to the y-niche, we anticipate that pattern on the x-axis should be likely to the extent that the x-niche dominates competitive interactions---i.e. when the process noise from the perspective of the x-niche is sufficiently low. To quantify this, we fix $w_x=0.1$, and test the x-axis for species clustering as we dial $w_y$ up or down. 

Species x- and y-niche values are drawn from a uniform distribution between 0 and 1, with the latter being redrawn for every new run. We simulate 100 replicates of each scenario to ensure that results are robust. In each replicate we tune the scaling factor $a_0$ to ensure that the community-wide average intensity of competition is the same in all our simulations. 

Lotka-Volterra equations approximate more complex dynamics near equilibrium \citep{Schoener1974, Schoener1976}, hence their widespread use in ecological theory. However, true consumer-resource interactions do not map trivially onto niche differences. We therefore tested our approach against models where species dynamics are explicitly based on their use of resources. Rather than prescribe the competition coefficients, we obtain them from simulation outcomes of the model, and then compare with distances in niche space. In Appendix B we present a resource-consumer model with two types of essential resources, such as different nutrients. The results confirm that the minimum niche distance is a good predictor of pairwise competitive impacts. In Appendix C we present a model of complementary niche axes: plants compete for space in a heterogeneous landscape, and high competition due to similar affinity for one environmental property (e.g. humidity) can be compensated by dissimilar affinities for another property (e.g. salinity). Results confirm that the Euclidean niche distance accurately predicts competitive interactions.

\subsection*{Measurement noise}
Even if niche space consists of a single niche axis, niche axes are typically not directly measurable; traits are. Whether clustering on a niche axis (e.g.\ competence in acquiring and processing light) translates as clustering on a measurable trait axis (e.g.\ maximum height) depends on how close is the relation between the trait and the niche axis (Fig.~\ref{Fig2}). We call the noise in this relation \emph{measurement noise}. 

To examine the impact of this type of noise, here we assume a simpler niche space than above, consisting of a single niche axis. The niche axis fully determines competitive interactions (i.e.\ there is no process noise), and we now assume that competition coefficients are monotonically tied to niche differences. This leads to strong clustering along the niche axis. 

Knowing that the niche axis is clustered, we want to assess the probability of clustering along a measurable trait axis, given that the trait is statistically associated with the niche. We then arrange species on a proxy trait axis such that trait values are random variables correlated with the niche values (Fig.~\ref{Fig5}A-D). Because of measurement noise, species relative positions on the niche axis will not be entirely preserved on the trait axis. As a result, competition will not be a strictly monotonic function of trait separation (Fig.~\ref{Fig5}E-H), and the clustered niche pattern will be somewhat randomized on the trait axis. Mathematically, we write $t_i=x_i+\varepsilon_i$, where $t_i$ is the trait value of species $i$, $x_i$ is its niche value, and $\varepsilon_i$ is the measurement noise between the two. The $\varepsilon_i$ are normally distributed random variables with mean 0 and whose variance determines the magnitude of the noise. Note that trait noise does not affect species dynamics, as the noise here is at the level of how we measure species and not at the level of their interactions. 

\subsection*{Using multiple traits to mitigate measurement noise}
In nature, a single niche axis will correlate with multiple trait axes. For example, niche strategy regarding competition for light in forests will reflect on specific leaf area, leaf nitrogen content, twig length, maximum plant height, etc. Individually, each trait will be an imperfect predictor of the species' light-related niche strategy, but together they may provide a better estimate. Combining multiple traits associated with the same niche axis may lead to a better estimate of the true niche values, and therefore clustering that is lost on any individual trait axis may appear on an aggregate axis. 

We test this idea using a simulated set of $n$ proxy traits, and their first principal component as the aggregate axis. The trait value $t^a_i$ of species $i$ on trait axis $a$ ($i=1,2,\dots,200;\;a=1,2,\dots,n$) is a random variable linearly related to its niche value $x_i$. We then write $t^a_i = b^a x_i + c^a + \varepsilon^a_i$, where $b^a$ and $c^a$ are constants specific to trait axis $a$ and $\varepsilon^a_i$ is the measurement noise, which is normally distributed with mean 0 and variance $\sigma^2_a$. This variance can be different for different trait axes, as the niche axis may be more tightly linked to some traits than others---e.g. the light-niche may be more directly influenced by maximum height than wood density. 

Although we write the trait as a function of the niche, we are not assuming the latter determines the former, but merely that they are correlated (see discussion in Appendix E). We also distinguish between multidimensional niche space and the present scenario of a one-dimensional niche axis that maps to multiple traits. An example of the former is competition based on the combined overlap in strategies for capturing light and strategies for acquiring water, whereas an example of the latter is when the light-niche strategy fully determines competitive interactions (possibly because all species involved have identical strategies in all other niche axes), and maps to various traits such as maximum height and specific leaf area. Finally, we note that our multiple traits correlate with each other through their association with the niche strategy, as is common in nature \citep{Litchman2008, Wright2010, Shoval2012}, but we are assuming no further correlation between these traits. This leads to the best-case-scenario where every new trait adds maximal information about the niche; were there any further correlations between the traits, new traits added to the analysis could be partially predicted by the existing ones, and therefore we would gain less information by incorporating them. 

\subsection*{Quantifying the noise}
Without process or measurement noise, competition will be strong between similar species, and weak between dissimilar ones. We therefore quantify noise as the degree of departure from this ideal scenario. We do so using  two summary statistics. 

Spearman's rank correlation coefficient, $\rho(A,d)$, quantifies the degree to which competition coefficients are monotonically related to pairwise distances on the niche or trait axis \citep{McDonald2009}. The index ranges from $-1$ (perfectly monotonic relationship) to 0 (no relationship).

As noted above, the kernel has a core (competition coefficients of species with small niche or trait differences) and a tail (large differences). We quantify the degree to which this core-tail structure predicts competitive relations by calculating the difference in proportion of tail and core elements that are greater than the kernel mean, $\pi(A) = P(A_\textrm{tail}>\overline{A}) - P(A_\textrm{core}>\overline{A})$. In the noise-free scenario, the first proportion is 0 and the second is 1, leading to a difference of $-1$, while in a completely random matrix both proportions are the same and the expected difference is 0. 

As these metrics put both process and measurement noise in the same footing as degrees of kernel disorder, they allow comparison of the relative impact of each.

\subsection*{Quantifying clustering}
To determine whether or not a species assemblage is clustered, we use a metric based on the k-means clustering algorithm and the gap statistic method \citep{MacQueen1967, Tibshirani2001}. The metric takes into account the niche or trait values of each species as well as their abundances, and estimates the number of clusters that best fits our species assemblage by comparing each fit against randomized null assemblages. The fit itself is based on pairwise distances along the niche or trait axis between all species in the community, weighted by their abundances (we represent each species by its average niche/trait value). Upon comparing results between the assemblage of interest and the null assemblages, we obtain a z-score and a p-value, which tell us the degree to which the assemblage is clustered, and whether the result is statistically significant. We provide the mathematical details of the metric in Appendix F.

\section*{Results}

Briefly, we found that the kernel's core-tail structure is key for pattern emergence, and the clustering pattern is robust to a substantial amount of noise between competition and niche strategies (process noise). On the other hand, clustering is sensitive to noise between niche strategies and measured trait values (measurement noise); this is because measurement noise effectively reshuffles species along the axis, thus randomizing the community. Finally, a large set of functional traits may help circumvent the issue, as combined they serve as a more accurate window into the biological niche than an individual trait is likely to be. 

\begin{figure}[H]
\caption{\textbf{Process noise}. \textbf{A, D}: Competition coefficients between a focal species and all other species, plotted against the difference in x-niche values. Although competition strictly declines with distances in the xy-niche plane, the relationship with x-niche differences alone is noisy. In the complementary niche axes case (A), most of that noise is between similar species---i.e. the core of the kernel. Conversely, in the essential niche axes case (D), most of the noise is between distinct species---i.e. the tail of the kernel. \textbf{B, E}: Simulation outcomes plotted in the xy-niche plane. Each disk represents a species, with disk size proportional to species abundance. In both the complementary niche axes case (B) and the essential niche axes case (E), there is strong clustering, as evidenced by the standardized magnitude of the clustering metric (Z in the legend) and the significant p-value (P). \textbf{C, F}: Same outcomes plotted on the x-niche axis only, with loss of the y-niche. Clustering is still visible, although the number of clusters is smaller because some distinct clusters cannot be resolved without the y-dimension. Notice also the lower magnitude of the clustering metric.}\label{Fig3}
\includegraphics[width=1\textwidth,angle=0]{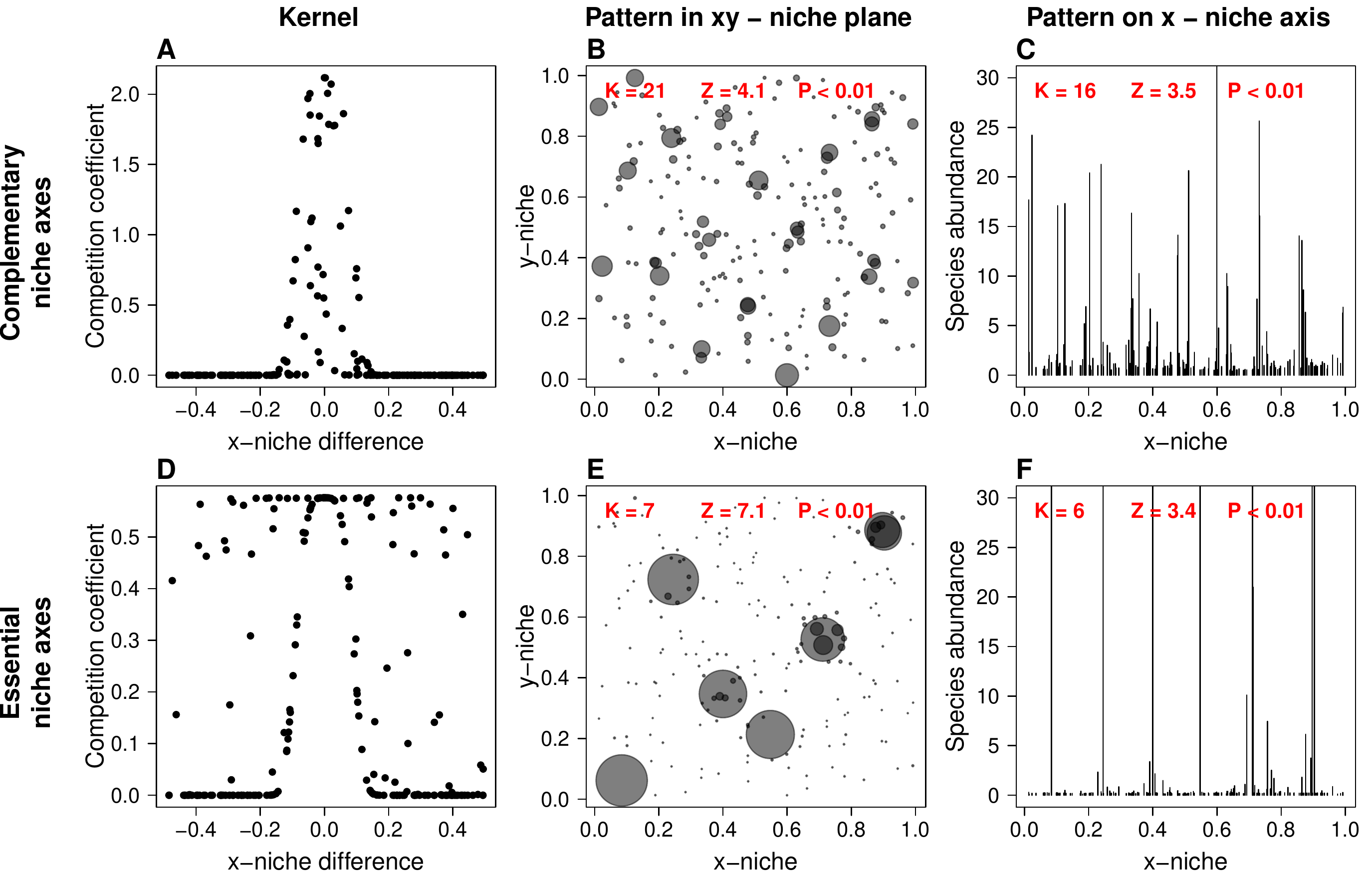}
\centering
\end{figure}

\subsection*{Process noise}

Fig.~\ref{Fig3} shows an example of each kind of process noise. In the complementary niche axes case, the noise is concentrated in the core of the kernel (Fig.~\ref{Fig3}A). In other words, competition is consistently low at large x-niche distances, but noisy at small distances. This is because some species pairs at short x-niche distances are far apart on the y-niche axis, thus bringing down the competition coefficient which would otherwise be high. Our clustering metric confirms that species are distinctly clustered in the xy-niche plane (Fig.~\ref{Fig3}B), as expected. However, even when the y-niche dimension is lost, clustering is still visible on the x-niche axis (Fig.~\ref{Fig3}C). Some distinct clusters collapse onto each other with the loss of the y-dimension, leading to detection of a lower number of clusters, but the general clumpy structure is preserved despite the noise. 

The essential niche axes scenario has the opposite type of noisy kernel (Fig.~\ref{Fig3}D): competition between species in the core is reliably high, while the tail is noisy because some species pairs that are far apart on the x-axis are close on the y-axis, bringing competition up. Again the xy-niche space is strongly clustered as expected (Fig.~\ref{Fig3}E), and again collapsing the y-dimension does not erase the clustering pattern, which is still visible from the x-niche axis alone (Fig.~\ref{Fig3}F). 

\begin{figure}[H]
\caption{\textbf{Probability of clustering by magnitude of noise}. The probability of clustering is estimated as the percentage of runs of each noise scenario where clustering was statistically significant ($p\le 0.05$), out of a total 100 replicates each. The noise is measured as the degree of randomness between the kernel and the x-niche axis (process noise) or the trait axis (measurement noise), using two indices: \textbf{A}: Monotonic decline of competition with species differences, measured with Spearman's rank correlation coefficient $\rho$. \textbf{B}: Core-tail structure, defined as the difference between the proportion of tail and core elements that exceed the kernel mean. Scenarios shown are process noise with complementary niche axes (blue curves), process noise with essential niche axes (red), and measurement noise (black). Compare with Fig.~S4.} \label{Fig4}
\includegraphics[width=1\textwidth,angle=0]{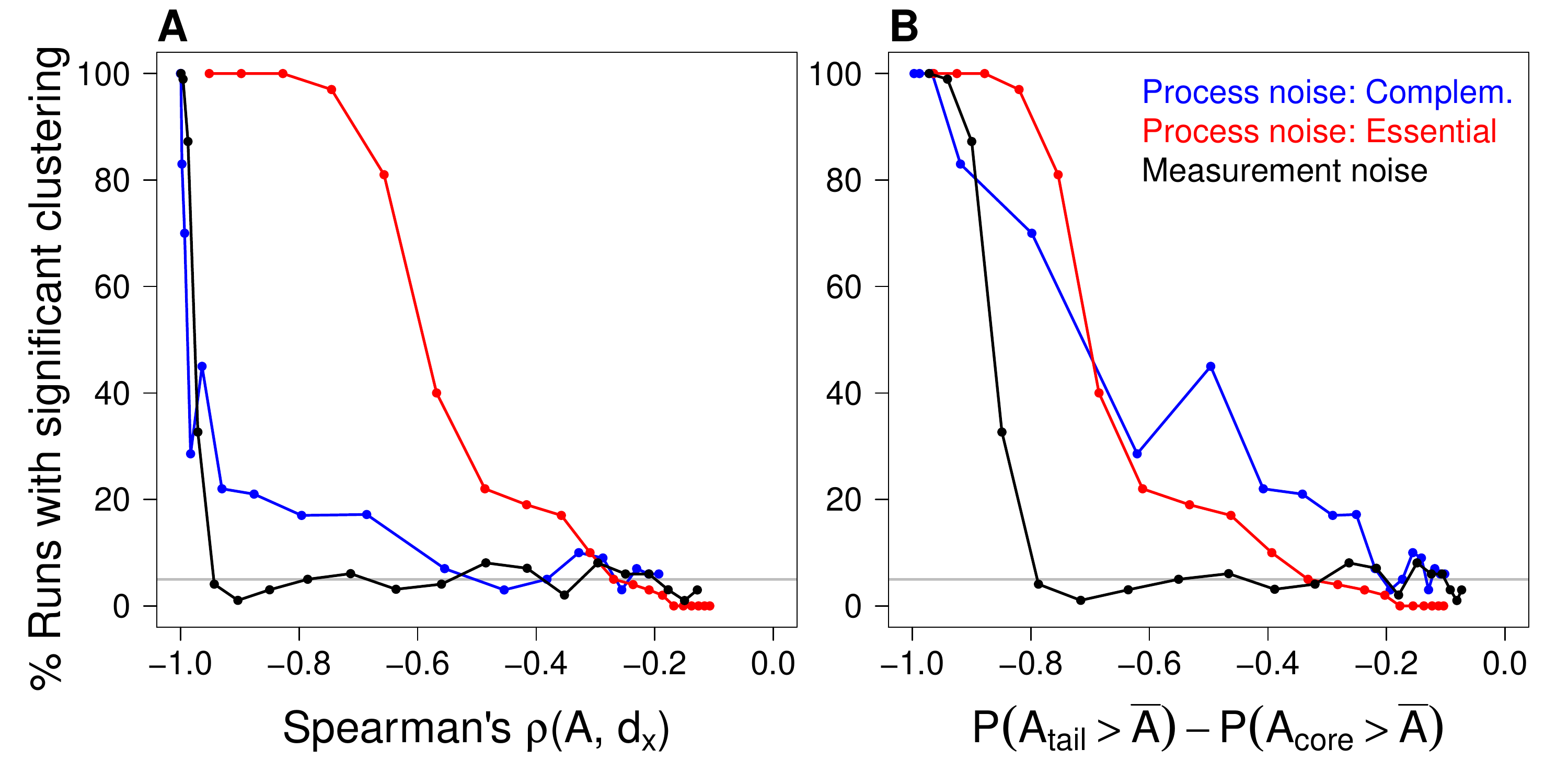}
\centering
\end{figure}

In both of these examples, the process noise onto the x-niche produced by the unacknowledged y-niche contributions is not sufficient to dissipate the pattern on the x-niche axis. However, as the contribution of the y-niche to the competitive relations increases, so does the process noise between the competition coefficients and distances on the x-niche axis, and the clustering on the x-axis weakens and eventually disappears (Fig.~\ref{Fig4}).

As the rank correlation between competition coefficients and niche differences degrades from $\rho=-1$ (fully monotonic) to $\rho=0$ (no correlation), the probability of clustering declines (Fig.~\ref{Fig4}A). In the case of complementary niche axes the drop is fast, but only reaches background levels near $\rho=-0.5$. With essential resource axes, clustering remains very likely until $\rho\approx -0.5$, at which point it quickly falls off, but only drops to background levels near $\rho=-0.2$.

The strength of the core-tail structure in the kernel predicts clustering in the face of process noise reasonably well, whether the niche axes are complementary or essential (Fig.~\ref{Fig4}B). In both scenarios, that probability drops below 50\% when $\pi\approx -0.7$ (Fig.~\ref{Fig4}B). 

In Appendix D we present a generalized process-noise model where we assume an indeterminate number of niche axes and other contributing factors adding noise to the relationship between competition and the known niche axis. We find similar results as the two-dimensional niche space model, indicating that higher niche dimensions add no qualitatively new phenomena.

These results indicate that even when niche space is multidimensional and part of it is unknown, we can expect clustering along a single niche axis to the extent that, on average, species with high niche similarity compete more than dissimilar species. A fully monotonic relationship between competition and niche difference is not strictly required. 

\begin{figure}[H]
\caption{\textbf{Measurement noise}. \textbf{A-D}: Species niche values plotted against their proxy trait values. Measurement noise increases from A to D, leading to progressively lower coefficient of determination $R^2$ between niche and trait values. \textbf{E-H}: Competition coefficients between a focal species near the center of the niche axis (marked by the red dot in A-D) and all other species, plotted against trait difference between them. When niche and trait values are perfectly correlated, the kernel decreases monotonically with absolute trait differences (E); as the niche-trait connection loosens, the kernel becomes noisy (F to H). Legend numbers show the noise scores according to each of our two statistics, Spearman's $\rho(A,d)$, and $\pi(A)=P(A_{\textrm{tail}}>\overline{A})-P(A_{\textrm{core}}>\overline{A})$. \textbf{I-L}: Species abundance plotted against trait values. Distinct clusters are visible when niche and traits match (I), but the pattern quickly fades as $R^2$ drops below 1 (J to L). Legend shows the estimated number of clusters and the z-score of the clustering statistic. Stars indicate statistical significance ($^{**}: p\le 0.01$;  $^*: p\le 0.05$).}\label{Fig5}
\includegraphics[width=1\textwidth,angle=0]{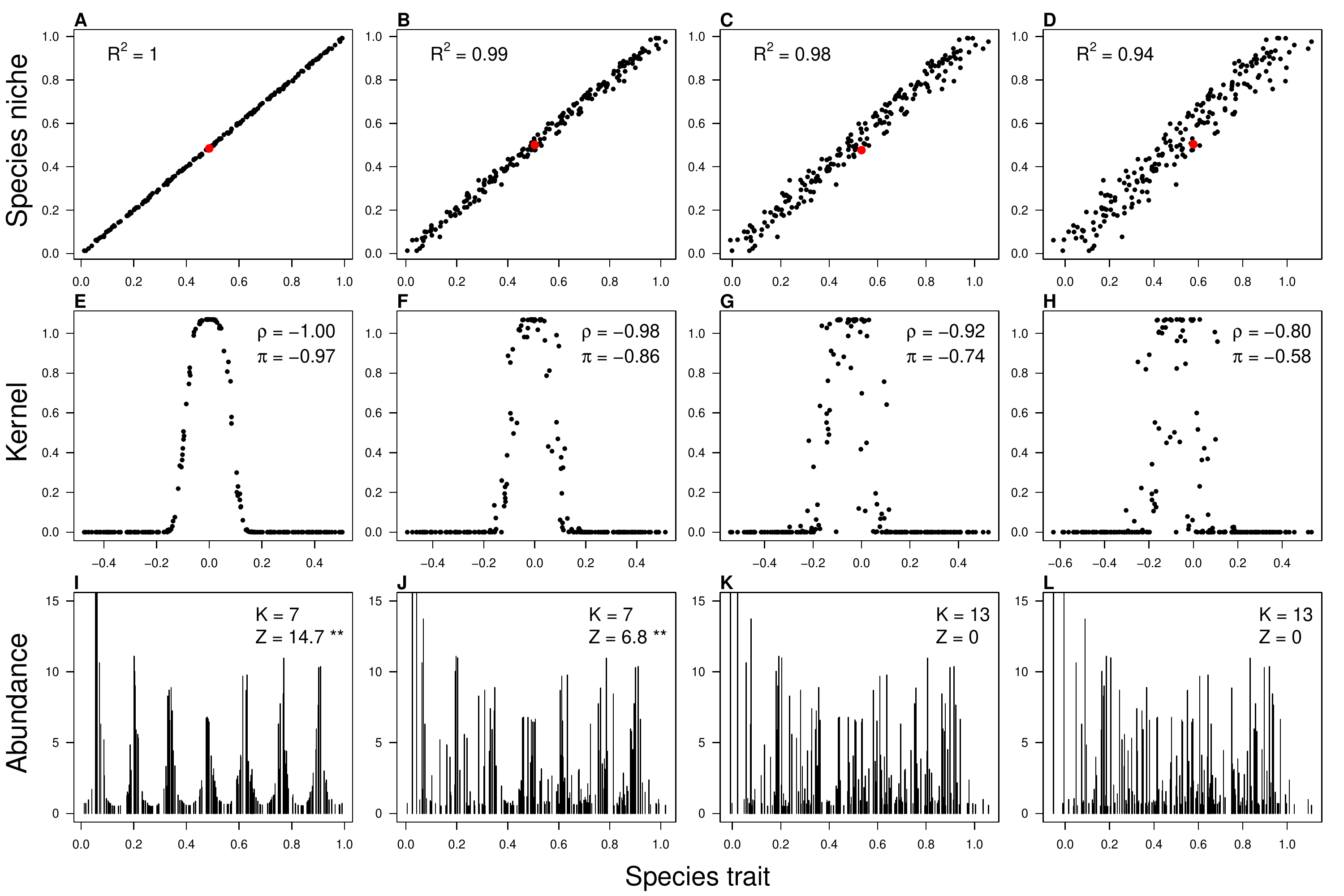}
\centering
\end{figure}

\subsection*{Measurement noise}
When the noise is between niche strategies and measured trait values (measurement noise), the probability of clustering on the trait axis quickly drops to zero; faster in fact than with noise between competition and niche strategies (process noise), as one can see by comparing the black curve to the colored curves in  Fig.~\ref{Fig4}. A glance at the correspondence between niche and trait values (Fig.~\ref{Fig5}A-D), compared with the pattern on the trait axis (Fig.~\ref{Fig5}I-L), shows that the pattern quickly fades to oblivion even while the coefficient of determination between niche and trait values is still very high ($R^2>0.9$). This occurs not because species are not sorting into clusters, but because our imperfect access to their true niche values misaligns them on the x-axis (Fig.~\ref{Fig5}I-L). In doing so, measurement noise effectively brings the species assemblage closer to the null model against which we compare it. Not only does statistical significance vanish, but the clustering metric also fails to recognize the true number of clusters (compare legend on Fig.~\ref{Fig5}I-L). 

This bodes ill for the prospects of finding niche-assembly patterns in nature, as the exact aspects of phenotype responsible for interactions are unlikely to be known or knowable, and proxy traits will often be our best resource to estimate them (Appendix A).

We note that the plot of competition coefficients against trait differences (as opposed to niche differences) shows visible core-tail structure (Fig.~\ref{Fig5}E-H). Indeed, it resembles process noise with complementary niche axes (cf.~Fig.~\ref{Fig3}A). And yet, the corresponding patterns are very different. A closer look reveals that the measurement-noise kernel is more structured than the core-noise kernel (Appendix G): unlike the process noise scenario, the matrix is symmetric and the noise is autocorrelated. In fact, the measurement-noise kernel contains the same set of elements as the noise-free kernel, and one can recover the latter via a permutation of the rows and columns of the former. We believe this non-random structure relative to the core-noise scenario is key to understanding the difference in the respective patterns, but at present we lack mathematical proof. 

\subsection*{Using multiple traits to estimate niche values}

\begin{figure}[H]
\caption{\textbf{Using multiple traits to mitigate measurement noise}. \textbf{A}: Degree of clustering on a first principal component axis increases with the number of independent traits used in the principal component analysis (PCA). Curves represent scenarios with different degree of measurement noise, with an average $R^2$ between trait and niche values across species ranging from 0.5 (lightest shade of gray) to 0.9 (darkest), at intervals of approximately 0.05. Values shown are averaged across 100 runs of each scenario. Red line shows the clustering z-score on the niche axis. \textbf{B}: Lowest number of traits required to detect significant clustering on the first component axis, as a function of the average $R^2$ between trait and niche values. Black and red curves show results using significance threshold $\alpha=0.05$ and 0.1, respectively. Values shown are averaged across 100 runs of each scenario.}\label{Fig6}
\includegraphics[width=1\textwidth,angle=0]{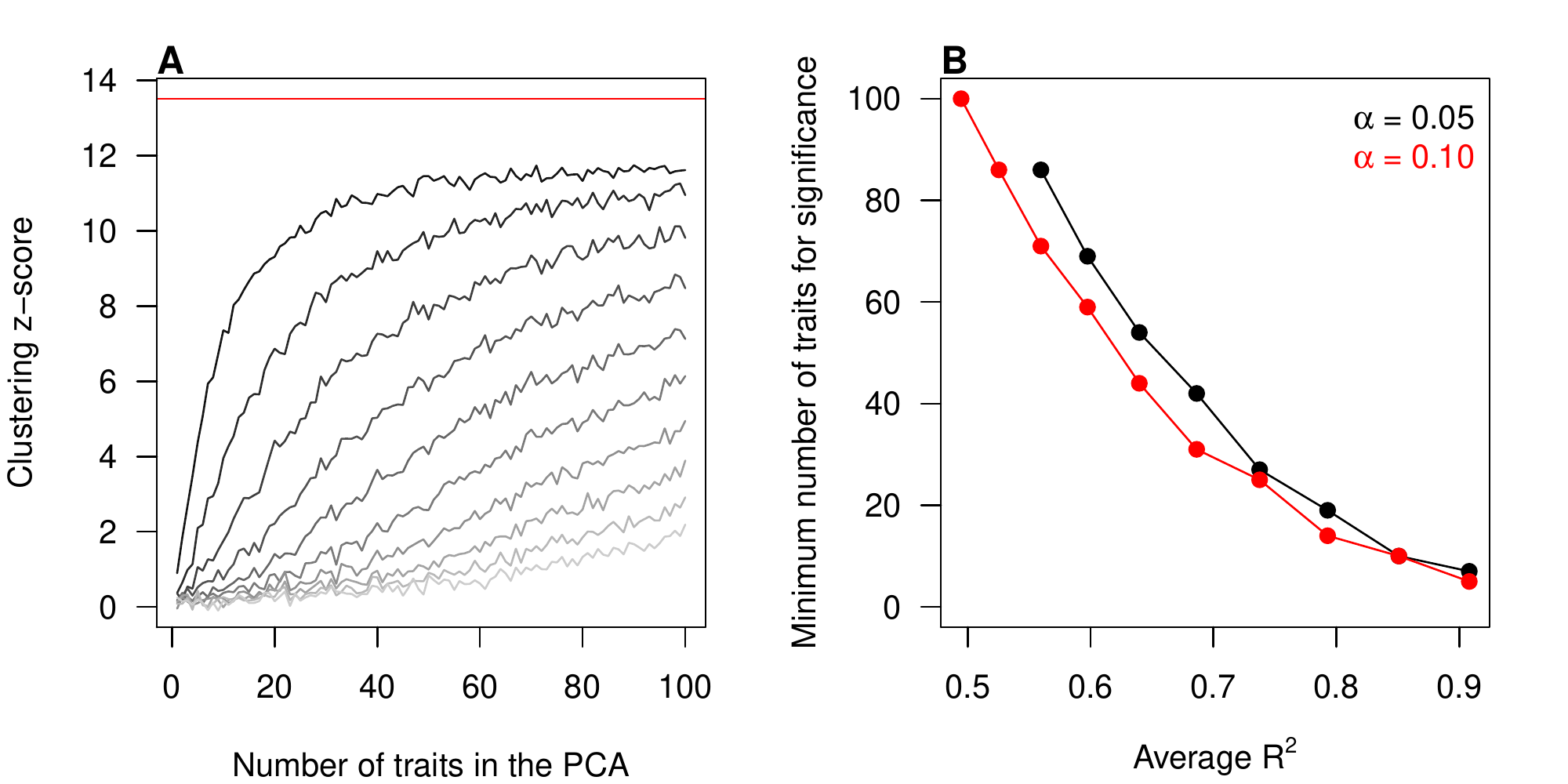}
\centering
\end{figure}

When multiple traits are related to the niche axis and are otherwise uncorrelated with each other, their first principal component is a better proxy for the niche axis than most of the individual traits. Other things being equal, clustering on the first component is stronger as more traits are used (Fig.~\ref{Fig6}A), reflecting the fact that the quality of the first component as a proxy for the niche axis increases with the number of traits measured. However, the number of traits required for statistically significant clustering on the first component can be very high. Even when the average $R^2$ between niche and trait values is as high as 0.9, we needed at least 5 traits for significant clustering at the $p\leq 0.1$ level and 7 traits at the $p\leq 0.05$ level. For lower $R^2$, denoting looser ties between trait and niche levels, that minimum number quickly rises (Fig.~\ref{Fig6}B).

We note that although the first component is typically a better proxy than an individual trait axis, it will not necessarily be better than every single trait (see Fig.~S5 in Appendix E). If one knows that a given trait is a much better proxy for species niche values than all other traits, it may be better to use that trait alone rather than to run principal component analysis on a set of traits. Realistically, however, such cases would be exceptional, as ecologists have only imperfect knowledge of the connection between measurable traits and niche strategies. Compared to traits with a moderate to high association with the niche axis, the first principal component has a higher chance of revealing niche pattern (Fig.~S5); and the more traits, the higher that advantage. Of course, no number of traits is a substitute for high-quality traits, i.e. traits demonstrated to reflect species niche strategies. This is especially true given the high sensitivity of pattern to measurement noise, as seen above.

In general, the number of traits required for detecting pattern will depend on their relation with the niche, which may be more complicated than presented here. Our linear model illustrates how even in a simple case, many traits may be necessary to parse the pattern.

\section*{Discussion}

Understanding how competition drives biodiversity patterns remains a challenge in ecology. The hypothesis of self-organized similarity or emergent neutrality \citep{Scheffer2006, Holt2006} states that competing species may sort into groups of species with similar traits. This concept may generalize classical limiting similarity theory \citep{DAndrea2016b} and reconcile it with contrasting empirical findings \citep{Chase2003, Gotzenberger2012}. However, the relevance of these ideas for field ecologists is limited by assumptions idealizing the connection between competition and niche strategies, and between niche strategies and measurable traits. Here we tested how uncertainty in these connections impacts the prospects of detecting the pattern in nature. 

When we introduced noise in the link between competition and niches, we found considerable robustness. Clustering remains likely as long as the kernel has a distinctive core-tail structure. In other words, competition between species far apart on the niche axis must be on average lower than between those close together. That the required trend is statistical rather than strict makes clustering on a single niche axis a distinct possibility in nature, even when competition is the product of multiple niche axes. Of course, probing multiple niche axes will make any pattern even clearer as it should minimize process noise. However, this is not automatically achieved by measuring multiple traits. Trait space is constrained by traits' roles in ecological performance \citep{Shoval2012}, and in fact a large set of traits may collapse into a single niche axis \citep{Wright2004}. To properly assess multidimensional niche space, one must use traits related to different niche dimensions. Knowledge of these dimensions and their relationship to traits requires natural history expertise. Absent that, process noise is inevitable. 

In contrast, we found high sensitivity to measurement noise. This is because measurement noise randomizes species positions along the true niche axis. In doing so, it brings the data closer to the null model. One cannot connect the distribution of species along an axis to mechanism if that distribution is random. Therefore, to stand a chance at finding pattern one must keep measurement noise at a minimum. This requires a tight link between niche axes and the actual traits measured in field studies, such that the proxy traits provide high-accuracy estimates of species niche strategies. A large set of traits with a demonstrated role in niche strategies is better than a few traits with weak functional connections. Demonstrating such roles requires explicit theory \citep{Pacala1998, Kohyama2006}, and extensive empirical work \citep{Violle2007, Litchman2008, Sterck2011, Herben2014}. 

The kernel contains all information about species interactions, and hence holds the key to pattern. Mathematical proofs connect kernel and trait patterns in simpler models lacking noise \citep{Hernandez-Garcia2009, Fort2009, Leimar2013}. However, more general kernels are less amenable to analytical proofs, and as yet have only been studied via simulations. Our results indicate that core-tail kernel structure is critical for trait pattern, but a more complete theory predicting the pattern from the shape of the kernel is still missing. Providing a low-dimensional representation of the kernel that predicts pattern in realistic noisy systems remains an open challenge. 

Our study is a step towards more realistic models of trait-based patterns of niche assembly. Our results complement previous explorations of trait-based niche patterns that considered other forms of noise such as demographic or environmental stochasticity \citep{Tilman2004, Gravel2006, Ernebjerg2011, DAndrea2018}. Attempts at higher realism would also require dedicated multidimensional niche models. Multidimensional niche space, despite having long been recognized as part and parcel of niche theory \citep{May1975, Tilman1982, Holt1987}, remains understudied in theoretical ecology (but see \citealt{Fort2010, Geange2011, Eklof2013}), and essentially untouched in the trait pattern literature. Finally, our models neglect non-competitive interactions, and require restorative forces such as immigration to replenish otherwise transient clusters. However, our results provide insight into the prospects of finding clusters in systems where both competition and dispersal are known to play a large role in community assembly, such as grasslands \citep{Tilman1994} and tropical forests \citep{Vellend2016}.

To our knowledge, we are the first to examine how uncertainty about the drivers of species interactions impacts ecologists' ability to detect their effect on community structure. Our findings highlight that theoretical results are only useful to field work when they are robust to uncertainty. More broadly, our work can initiate a more general picture of this robustness or lack thereof across different patterns and processes.

\section*{Acknowledgments}
J.OD. acknowledges the Simons Foundation Grant \#376199 and McDonnell Foundation Grant \#220020439. A.O. acknowledges the National Science Foundation Grant \#1038678, “Niche versus neutral structure in populations  and  communities”  funded  by  the  Advancing  Theory  in Biology program, the Danish National Research Foundation Grant \#DNRF96 through support of the Center of Macroecology, Evolution and Climate, and the Miller Institute for Basic Research in Science at the University of California, Berkeley.


\end{document}


\title{Appendices for Translucent windows: how uncertainty in competitive interactions impacts detection of community pattern}
\author{Rafael D'Andrea$^{1, \ast}$, Annette Ostling$^{2, \ast\ast}$ \& James O'Dwyer$^{1, \dagger}$ \\ 
  $^1${\small Dept.~Plant Biology, University of Illinois at Urbana-Champaign, IL, USA}\\
  $^2${\small Dept.~Ecology \& Evolutionary Biology, University of Michigan, Ann Arbor, MI, USA}\\  $^{\ast}${\small \texttt{rdandrea@illinois.edu}}, $^{\ast\ast}${\small \texttt{aostling@umich.edu}}, $^{\dagger}${\small \texttt{jodwyer@illinois.edu}} 
}
\date{}

\maketitle

\appendix
\renewcommand\thefigure{S\arabic{figure}} 
\setcounter{figure}{0}

\section{Niche space}\label{appBox}
Clusters emerge in models where species belong in a niche space, typically a one-dimensional line, and their competitive impact on one another is a function of their mutual distances in that space. The definition of niche space here is abstract and operational: it is the set of aspects of organism performance that affect competition; species sufficiently far apart in that space can stably coexist.

For example, in a model of birds competing for two substitutable food types, say pollen and insects, a species' niche is the amount of time it spends foraging for each food. Assuming the total foraging time is fixed, the niche space is a one-dimensional axis, with points on that axis indicating how long the species spend on one of the resources---the time spent on the other food is then set, as both times need to add to a fixed amount. If there are three substitutable resources, say pollen, insects, and seeds, then niche space is a two-dimensional plane, and so on. The classical MacArthur-Levins (1967) model assumes a continuous resource axis with an indefinite number of food types---say, seeds falling on a size continuum---which in principle makes niche space infinite-dimensional. However, the number of dimensions drops if there are further constraints on the amount of time a bird can spend on each food. \citet{MacArthur1967} considered the extreme scenario where the relative time spent on one of the resources fixes the time spent on all others: they assumed every species share a resource utilization curve that differs only by what resource it peaks on, i.e. the seed they eat the most. In that case, niche space collapses onto a single axis.

Under this operational definition, a species' niche is not necessarily a resource, measurable trait, or the environment where it lives. Those are only associated with the niche to the extent that they affect competition. In the example of birds competing for seeds falling on a size continuum, the assumption of same-shape resource utilization curves across all species means a species' favorite resource suffices to determine its niche. In this case the niche itself can be equated with the favorite seed size. More generally the niche would be a list of numbers reflecting the amount of time spent on seeds of each size. A relevant trait such as beak size will correlate with its favorite food \citep{Schluter1984}, but even in this idealized example of a one-dimensional niche axis the correlation between niche and trait need not be one-to-one, as birds' beaks are also shaped by other functions such as singing \citep{Huber2006}.

\section{Competition for essential resources leads to tail noise, and minimum distances in niche space are a good predictor of competitive interactions}\label{appResCons}

Consider a resource-consumer model with two resource axes. Resources grow logistically in the absence of consumers. Consumers’ niche strategies determine their resource preference along each resource axis, but they need resources from both axes to grow. One example is animals that thrive on a diet containing both plant and animal foods. We implement this by having consumer growth rates depend multiplicatively on resources from both axes \citep{Saito2008}:
\begin{equation}
\frac{\ud A}{\ud t}=\mu-m A + \left(\sum_{i=1}^R e_x \gamma_{AX_i} X_i \sum_{i=1}^R e_y \gamma_{AY_i} Y_i \right)A                  
\end{equation}

where A is the abundance of the consumer, m is its intrinsic mortality, $\mu$ is the immigration rate, R is the number of resources on each axis, $X_i$ are the abundances of the x-resources, $\gamma_{AX_i}$ quantifies the consumer preference for the i-th x-resource, and $e_x$ is the conversion efficiency---similarly for the y resources. The product in the consumption derives from the law of mass action in chemistry, in this case representing a reaction that needs three components to occur (the consumer and the two resource types). In our biological context, the term reflects a situation where each resource type offers nutrients not found in the other type, and the consumption of both types are linked together because the consumer will lack the energy to forage if it lacks either. 

The corresponding resource equations are
\begin{align}
\frac{\ud X_i}{\ud t}&=\nu+X_i r\left(1-\frac{X_i}{K}\right)-\left(\sum_{N=1}^S \gamma_{NX_i} N \sum_{j=1}^S \gamma_{NY_j} Y_j \right) X_i\\            
\frac{\ud Y_i}{\ud t}&=\nu+Y_i r\left(1-\frac{Y_i}{K}\right)-\left(\sum_{N=1}^S \gamma_{NY_i} N \sum_{j=1}^S \gamma_{NX_j} X_j \right) Y_i
\end{align}
where $r$ is the resource intrinsic growth rate, $K$ the carrying capacity, $\nu$ the resource immigration rate, and $S$ the number of consumers. Notice that these equations reflect the abovementioned linked consumption of both resource types. 

We note that this type of resource-consumer interaction is not the only way to model colimitation by more than resource; another common approach is Liebig’s law of the minimum\footnote{Our multiplicative model corresponds to chemicals in a vat where the reaction is C + A + B $\rightarrow$ C + C, i.e. both substances A and B must be present at the same time for C to generate a new C; in contrast, the Liebig scenario corresponds to a reaction like C + A $\rightarrow$ CA; CA + B  $\rightarrow$ C + C, i.e. where there is a stable intermediate state such that C does not need to encounter A and B at the same time, but still needs both to generate a new C. In biological terms, Liebig's law corresponds to a case where the consumer acquires each resource independently of the other but will not grow without both, whereas in the multiplicative model the consumer's ability to acquire one of the resources depends on having access to the other.} \citep{Saito2008}. Empirically, consumers show a variety of responses to addition of essential resources, compatible with different models \citep{Harpole2011}. Our model predicts that consumers will respond to addition of either resource in isolation, and will respond maximally to addition of both resources. This has been observed in communities of bacteria \citep{Danger2008}, plankton \citep{Harpole2011}, and plants \citep{Craine2010}.

For simplicity, we set $\mu,\nu,r,K,e_x,e_y$ identical across species and resources, so that we can focus on the competitive effect of differences in resources preferences. We set the preference of consumer $A$ for resource $x_i$ to be a declining function of the difference between the resource and the consumer's favorite resource $x_A$ along that axis, $\gamma_{AX_i}=\exp(-((x_A-x_i)/w_x)^4)$. Similarly for y-resources.

\subsection*{Competition kernel}
Consumer-resource interactions do not map trivially to Lotka-Volterra competition coefficients that depend only on trait distances. In general, this mapping might only be valid in the vicinity of an equilibrium \citep{Schoener1974, Schoener1976}. In fact, expansions of small perturbations around equilibria is the traditional approach to converting resource-consumer models to Lotka-Volterra form, and thus extracting competition coefficients valid in that regime \citep{MacArthur1972,Abrams1980}. This is the approach we adopt here.

The competition kernel, which quantifies the competitive impact of consumers on one another, is comprised of the matrix elements $\alpha_{AB}= -\frac{\partial}{\partial B} \left(\frac{1}{A} \frac{\ud A}{\ud t}\right)\bigg|_{A^* B^* X^* Y^*}$, where the asterisk denotes equilibrium abundances. We calculate it explicitly via linearization around the equilibrium:
\begin{equation}
A(t)=A^*+a(t); \;\; X_i (t)=X_i^*+x_i (t); \;\; Y_j (t)=Y_j^*+y_j (t) \nonumber
\end{equation}
Substituting in the consumer equation and discarding higher order terms:
\begin{align}
\frac{\ud A}{\ud t}&=\mu-(A^*+a)m+(A^*+a) e_x e_y \sum_i\sum_j \gamma_{AX_i}\gamma_{AY_j} (X_i^*+x_i)(Y_j^*+y_j)\nonumber\\
&\approx -a\frac{\mu}{A^*}+A^* e_x e_y \sum_{ij} \gamma_{AX_i}\gamma_{AY_j} (Y_j^* x_i+X_i^* y_j )\nonumber
\end{align}
The competition coefficients are then
\begin{align}
\alpha_{AB}&=-\frac{\partial}{\partial B} \left(\frac{1}{A} \frac{\ud A}{\ud t}\right)\nonumber\\
&=\delta_{AB}\frac{\mu}{A^{*2}}-e_x e_y \sum_{ij} \gamma_{AX_i}\gamma_{AY_j} \left(Y_j^* \frac{\partial x_i}{\partial B}+X_i^*\frac{\partial y_i}{\partial B}\right)\nonumber\\
&=\delta_{AB}\frac{\mu}{A^{*2}}+e_x e_y \left((A\textbf{Y}^*)\cdot\left(-\frac{\partial \textbf{x}_i}{\partial B}\right)+(A^T \textbf{X}^*)\cdot\left(-\frac{\partial \textbf{y}_i}{\partial B}\right)\right)\label{eq4}
\end{align}
where in the last line we used compact matrix notation, with $A_{ij}\equiv \gamma_{AX_i}\gamma_{AY_j}$, and the dot indicating internal product between the vectors.

To proceed we must write $\textbf{x}$ and $\textbf{y}$ in terms of the consumer abundances. We do so by assuming resources undergo faster dynamics than consumers:
\begin{align}
0=\frac{\ud X_i}{\ud t}&=\nu+r(X_i^*+x_i )\left(1-\frac{X_i^*+x_i}{K}\right)-(X_i^*+x_i ) \sum_N\sum_j\gamma_{NX_i}\gamma_{NY_j}(N^*+n)(Y_j^*+y_j )\nonumber\\
&=-\frac{\nu}{X_i^*}x_i-rX_i^*\frac{x_i}{K}-X_i^* \sum_N\sum_j \gamma_{NX_i}\gamma_{NY_j}(N^* y_j+Y_j^* n)\nonumber\\
\rightarrow x_i&=-\frac{1}{\frac{\nu}{X^{*2}_i}+\frac{r}{K}}  \sum_N\sum_j \gamma_{NX_i}\gamma_{NY_j}(N^* y_j+Y_j^* n)\nonumber
\end{align}
We can write this expression in compact matrix notation:
\begin{equation}
\textbf{x}=P^x \textbf{y}+Q^x \textbf{Y}^*\nonumber
\end{equation}
where $P_{ij}^x=-\epsilon_i^x \sum_N N^* \gamma_{NX_i}\gamma_{NY_j}$ , $Q_{ij}^x=-\epsilon_i^x \sum_N n\gamma_{NX_i}\gamma_{NY_j}$, and $\epsilon_i^x=1/(\nu/X_i^{*2}+r/K)$. Notice that matrix $P^x$ is a constant whereas matrix $Q^x$ depends on the consumer perturbations.

We get an analogous expression for the y resources:
\begin{equation}
\textbf{y}=P^y \textbf{x}+Q^y \textbf{X}^*\nonumber
\end{equation}
with parameters mirroring those of the x resources. 

Solving this linear system for $\textbf{x}$ and $\textbf{y}$, we get
\begin{align}
\textbf{x}&=(\textbf{1}-P^x P^y )^{-1} (P^x Q^y \textbf{X}^*+Q^x \textbf{Y}^* )\nonumber\\
\textbf{y}&=(\textbf{1}-P^y P^x )^{-1} (P^y Q^x \textbf{Y}^*+Q^y \textbf{X}^* )\nonumber
\end{align}
Substituting these in Eq.~(\ref{eq4}) and defining $B_{ij}^x\equiv \epsilon_i^x\gamma_{BX_i}\gamma_{BY_j}$ and $B_{ij}^y\equiv \epsilon_i^y \gamma_{BY_i}\gamma_{BX_j}$, we obtain our competition kernel:
\begin{align}
\alpha_{AB}&=\delta_{AB} \frac{\mu}{A^{*2}}\nonumber\\
&+e_x e_y (A\textbf{Y}^*)\cdot((\textbf{1}-P^x P^y )^{-1} (P^x B^y \textbf{X}^*+B^x \textbf{Y}^*))\nonumber\\
&+e_x e_y (A^T \textbf{X}^*)\cdot((\textbf{1}-P^y P^x)^{-1} (P^y B^x \textbf{Y}^*+B^y \textbf{X}^*))\label{eq7}
\end{align}

\subsection*{Results}
First, we note that if we unplug the y resources from consumer dynamics, such that only the x resources regulate those species, then differences in x-niche strategies are almost perfect predictors of pairwise competitive relations. In other words, if niche space is one-dimensional, niche strategies fully determine competitive interactions, and we therefore have no process noise.

\begin{figure}[H]
\caption{\textbf{Generalized process noise}. \textbf{A}. Species abundances by their x-niche. Legend shows number of clusters (K), z-score of the gap statistic (Z), and corresponding p-value (P); \textbf{B}. Resource abundances along each resource axis; \textbf{C}. Pairwise competition coefficients calculated from consumer and resource abundances using Eq.~(\ref{eq7}), plotted against species differences in x-niche strategies. There is a visible negative correlation. Legend shows rank correlation (Spearman’s $\rho$); \textbf{D}. Competition coefficients plotted against the minimum between dx and dy (scaled by the width of the respective resource use functions $\gamma$). Parameters used in the simulation: $S=R=100;\ e_x=e_y=1;\ K=0.1;\ r=100;\ m=1;\ \mu={10}^{-5};\ \nu=0.1;\ w_x=0.11;\ w_y=0.08$. All species and resources start with identical abundances, and we run the simulation until abundances are no longer changing.} \label{FigS_EssentialResources}
\includegraphics[width=1\textwidth,angle=0]{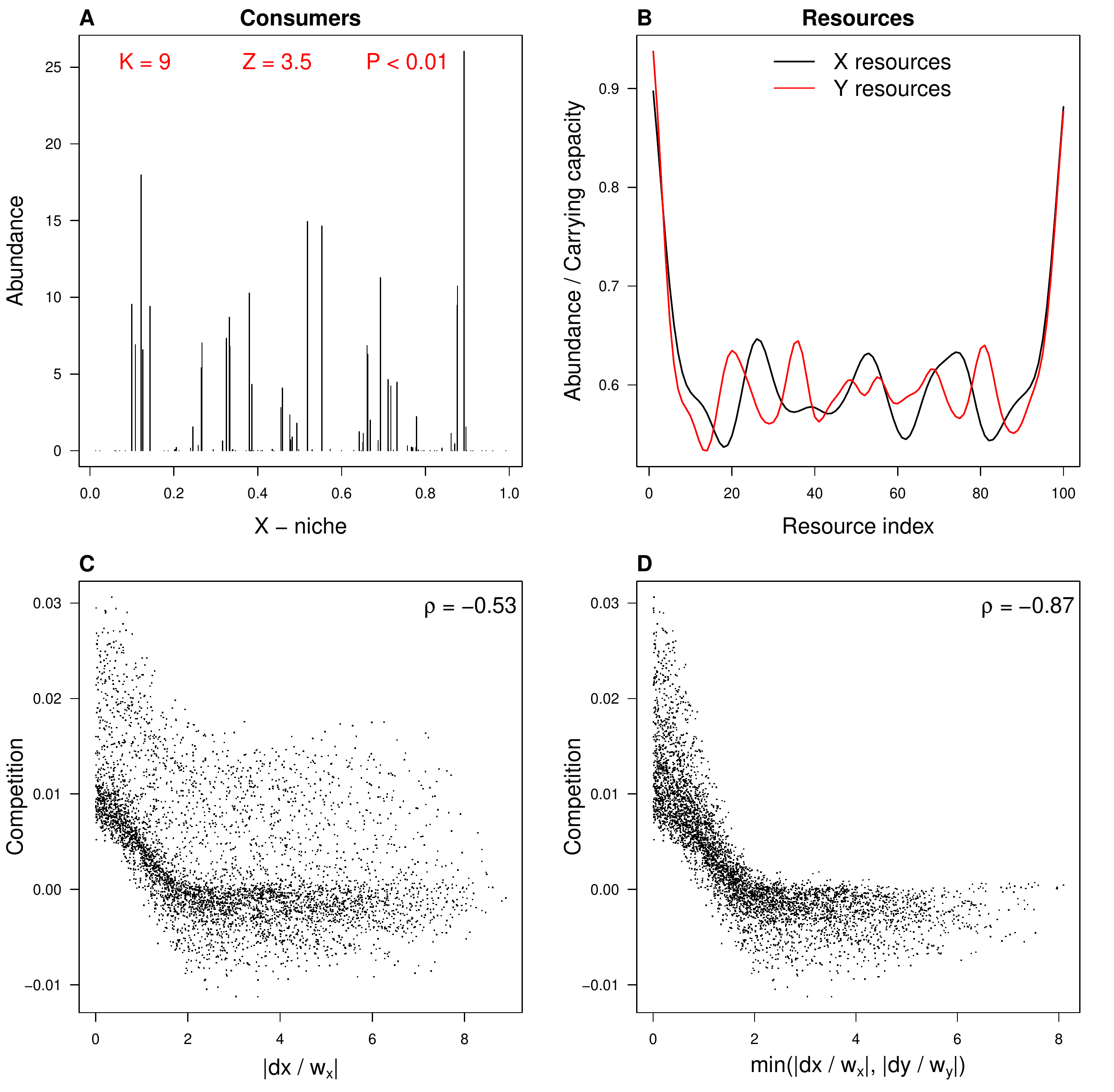}
\centering
\end{figure}

Next, we add the y resources back in. Fig.~\ref{FigS_EssentialResources}A shows a community obtained by numerical simulation of this model, with species arranged by their x-niche. Despite niche space being two-dimensional, with both x and y resources modulating species dynamics, the clustered abundance pattern is visible along the x-niche axis, as confirmed by our clustering metric (results summarized in the legend of the plot). Thus, even if the y-niche strategies of these species were entirely unknown, we would still be able to infer niche dynamics from the x-axis alone. The competition kernel, as calculated via Eq.~(\ref{eq7}), correlates with x-niche distances (Fig.~\ref{FigS_EssentialResources}C), but the relationship is noisy because of the contribution of the y-niche. The hidden y-niche adds process noise.

To determine whether the noise was higher in the core or tail of the kernel, we fitted a cubic spline to each row of the kernel---which for each species in the community describes the competitive impact of all other species---and compared the expected kernel values from the spline against the observed values. Species pairs were deemed to be in the core or tail based on whether they were below or above the niche distance at which the splined kernel is at the community-level mean. Using the chi-squared statistic $\chi_i^2=\sum_j(\alpha_{ij}-\hat{\alpha}_{ij})^2/\hat{\alpha}_{ij}$, we found that the noise is higher in the tail ($\chi_{\textrm{core}}^2=0.04, \chi_{\textrm{tail}}^2=0.07$). 

The minimum of the niche distances on either axis is a better predictor of the competition coefficients (Fig.~\ref{FigS_EssentialResources}D). In particular, the minimum distance is a much better predictor of the kernel’s tail than the x-distance: unlike the x-distance, a large minimum distance is a reliable sign of low competitive interactions. 

These results support our assumptions in the main text that essential resource axes may lead to noise in the tail of the kernel, and that the pairwise competitive relations are strongly associated with the minimum distance in the full niche space.

To summarize our findings regarding competition for essential resources: when niche space is one-dimensional, niche strategies fully determine competitive interactions—i.e. there is no process noise. When niche space is multidimensional, the map from competition to niche strategies is noisy, but the minimum distance in niche space is a good approximation. If we have no information on one of the niche dimensions, this process noise is compounded, particularly in the tail. 

\section{Competition for available space in two niche dimensions leads to core noise, and Euclidean distances are a good predictor of competitive interactions}\label{appAbiotic}

Consider a community of plants in a heterogeneous landscape. Plants compete for space, and the landscape is composed of patches, each of which can contain one individual. The local environment in each patch is characterized by two independent indices (for instance, soil humidity and salinity), and different species are optimally adapted to different suites of environmental conditions (i.e. different combinations of humidity and salinity). In this situation, the resource is available space with specific environmental conditions. Niche space is two-dimensional, in the sense that species niche strategies pertain to their affinities for these two independent environmental properties. 

We can model such a scenario as follows:
\begin{align}
\frac{\ud N_{ikl}}{\ud t}&=-mN_{ikl}+b\,T_{ikl}X_{kl}N_i\label{eqn6}\\
X_{kl}&=c-\sum_i N_{ikl}\label{eqn7}
\end{align}
where $N_{ikl}$ is the abundance of species $i$ in patches with humidity $k$ and salinity $l$, and $X_{kl}$ is the number of currently unoccupied patches with those conditions. The first term in Eq.~\ref{eqn6} represents death, which for simplicity we assume occurs at the same rate $m$ for all individuals in all patches. The second represents births, which are mediated by the species affinity for the local environmental conditions, $T_{ikl}$. We assume no dispersal limitation: all individuals of species $i$ contribute offspring equally to all patches (hence the factor $N_i$ in the birth term, as opposed to a weighted combination of $N_{ikl}$). The birth rate $b$ is the same for all species and patches, but varies over time depending on abundances and site occupancy, such that the total number of births balances the total number of deaths. This portrays a community with a seed bank that enables newly available sites to be quickly colonized, but where births are limited by available space, which is in turn provided by deaths. The number of available patches of humidity $k$ and salinity $l$ is equal to the total number of patches $c$ in the community with those conditions, minus the total occupancy in these sites. For simplicity, we assume $c$ is the same for all environmental conditions.

We can sum Eq.~\ref{eqn6} over the patches to find the dynamic equation for species abundances:
\begin{equation}
\sum_{kl}\frac{\ud N_{ikl}}{\ud t}=\frac{\ud N_i}{\ud t}=-mN_i+b\tau_iN_i\label{eqn8}
\end{equation}
where $\tau_i=\sum_{kl}T_{ikl}X_{kl}$. Note that in order to match total births and deaths, $b=m(\sum_iN_i)/(\sum_i\tau_iN_i)$. This dependence of the birth rate on deaths reflects offspring's need for available gaps to recruit, and provides a stabilizing mechanism that prevents populations from growing indefinitely.

Substituting Eq.~\ref{eqn8} in the usual definition of the competition coefficients, $\alpha_{ij}=-\frac{\partial}{\partial N_j}\left(\frac{1}{N_i}\frac{\ud N_i}{\ud t}\right)\bigg|_{\textbf{N}^*}$, we get 
\begin{equation}
\alpha_{ij}=-\left(\frac{\partial \tau_i}{\partial N_j}b+\tau_i\frac{\partial b}{\partial N_j}\right)\label{eqn9}
\end{equation}
Before we can calculate those derivatives, we must note that $\tau_i$ and $b$ are defined in terms of patch-specific subpopulations. Therefore we must convert $\partial/\partial N_j$ into $\partial/\partial N_{jkl}$. This conversion depends on how an increment to the population $N_j$ is distributed across the different patches. The natural assumption is that it will occur in proportion to the species affinity for the local environment in these patches, reflecting local filters. In other words, $dN_{jkl}=\frac{1}{\sum_{k'l'} T_{jk'l'}}T_{jkl}dN_j$. The proportionality constant ensures that $\sum_{kl}dN_{jkl}=dN_{j}$. It then follows that $\frac{\partial}{\partial N_j}=\frac{1}{\sum_{k'l'} T_{jk'l'}}\sum_{kl}T_{jkl}\frac{\partial}{\partial N_{jkl}}$. 

Applying the derivatives in Eq.~\ref{eqn9} and using Eq.~\ref{eqn7}, we obtain 
\begin{equation}
\alpha_{ij}=b\sum_{kl}T_{ikl}\frac{T_{jkl}}{\sum_{k'l'}T_{jk'l'}}-\frac{m}{\sum_{i'}N_{i'}\tau_{i'}}+\frac{m\sum_{i'}N_{i'}}{(\sum_{i'}N_{i'}\tau_{i'})^2}\left(\tau_j-\sum_{i'}N_{i'}\sum_{kl}T_{ikl}\frac{T_{jkl}}{\sum_{k'l'}T_{jk'l'}}\right)\label{eqn10}
\end{equation}
This expression is not particularly illuminating, but we note that it contains terms like $\sum_{kl}T_{ikl}T_{jkl}$, which represent the overlap in environmental preferences between species $i$ and $j$. This is in the same spirit as the classical work by MacArthur and others \citep{MacArthur1967, MacArthur1972}.

To complete the model, we must specify the environmental preferences of our different species. Suppose species $i$ has a preferred humidity level $x_i$ and salinity level $y_i$. Those preferences define the species niche strategy. Its affinity for a patch with humidity $x_k$ and salinity $y_l$ will then be
\[ 
T_{ikl}=\exp\left[-\left(\left|\frac{x_i-x_k}{w_x}\right|^2+\left|\frac{y_i-y_l}{w_y}\right|^2\right)^2\right]
\]
In words, the affinity is a declining function of the (two-dimensional) difference between the local environment and the species' preferred environment. The weighting factors $w_x$ and $w_y$ set the relative scales of the x- and y-environmental dimensions. For instance in the limit where $w_y\gg w_x$, species are indifferent to the y-dimension relative to the x-dimension. This will occur for example if they have similar tolerance to all salinity levels within the landscape's range. In that case, niche space is one-dimensional. 

\begin{figure}[H]
\caption{\textbf{Competition for available space in two niche dimensions leads to core noise}. \textbf{A}. Species abundances by their x-niche. Legend shows number of clusters (K), z-score of the gap statistic (Z), and corresponding p-value (P); \textbf{B}. Pairwise competition coefficients calculated from consumer abundances using Eq.~(\ref{eqn10}), plotted against species differences in x-niche strategies. There is a visible negative correlation, but it is noisy, especially in the core; \textbf{C}. Competition coefficients plotted against differences in y-niche strategies show a similar pattern. \textbf{D}. Plotting competition against the Euclidean distance (with dx and dy scaled by the respective weighting factors) resolves the core noise. Parameters used in the simulation: $S=100\ \textrm{species};\ R=10\ \textrm{different environmental levels in each dimension};\ m=0.01;\ w_x=0.1;\ w_y=0.38$. Both x- and y-niche strategies range from 0 to 1, as do environmental levels. All species start with identical abundances, totaling the carrying capacity of the community, such that community size does not vary. We then run the simulation until abundances are no longer changing.} \label{FigS_AbioticResources}
\includegraphics[width=1\textwidth,angle=0]{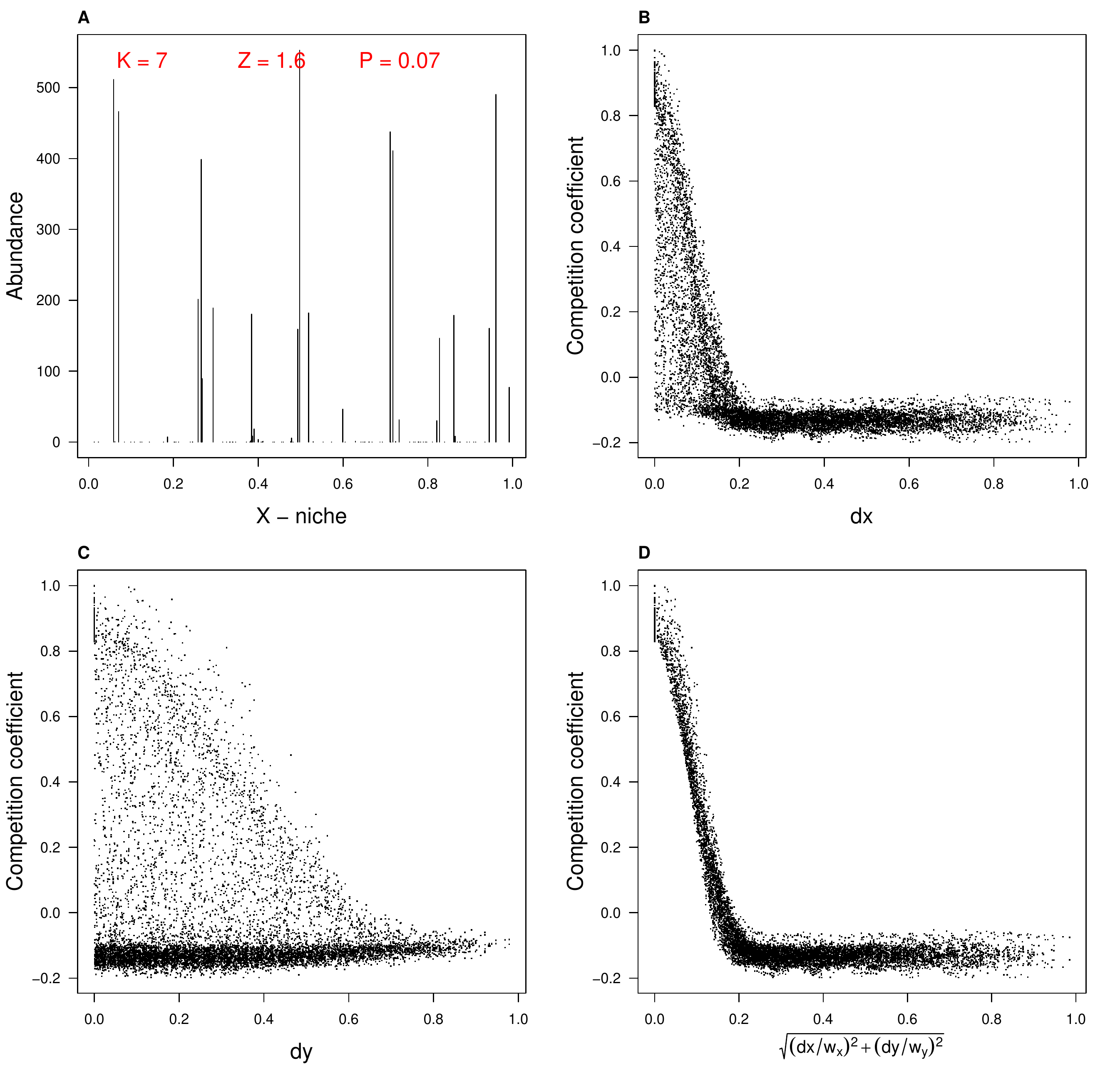}
\centering
\end{figure}

Results from numerical implementation of this model are shown in Fig.~\ref{FigS_AbioticResources}. Species cluster along their x-environmental preferences (the x-niche, Fig.~\ref{FigS_AbioticResources}A), even though competition does not correlate perfectly with differences in x-preference (Fig.~\ref{FigS_AbioticResources}B). The noise, visible as a scatter in the cloud of points, occurs mostly between species with similar x-preferences. In other words, the noise is greater in the core than in the tail. (Contrast this with the tail-heavy noise from the resource-consumer model of competition for essential resources, Appendix \ref{appResCons}.) The correlation with differences in y-preference is also imperfect (Fig.~\ref{FigS_AbioticResources}C). It is only when we plot competition coefficients against the Euclidean distance that we see a tight relationship (Fig.~\ref{FigS_AbioticResources}D). 

In summary, when we ignore either niche dimension, we face noise in the correspondence between competitive interactions and the observed niche strategies. That noise is concentrated in the core of the competition kernel. This is because if all we know is that two species are similar with respect to their x-preference, we may overestimate how strongly they compete because they may compensate for their x-similarity with very dissimilar y-niche requirements. On the other hand, the Euclidean distance in the two-dimensional niche space is an excellent predictor of competition. These results support our simple Lotka-Volterra model for complementary resources presented in the main text. 

\section{Generalized process noise}\label{appMultiNiche}
\begin{figure}[H]
\caption{\textbf{Generalized process noise}. \textbf{A-D}: Competition coefficients between a focal species near the center of the niche axis and all other species, plotted against niche difference between them. In the noise-free case (A), the kernel peaks at zero niche difference and decreases monotonically with absolute niche differences. In the noise scenarios the noise is concentrated in the core (B), tail (C), or uniform across the kernel (D). All kernels have the same average, $\overline{A}=S^{-2}\sum_{ij}A_{ij}=0.2$. Legend numbers show the noise scores according to each of our two metrics, namely Spearman's $\rho$ and $\pi(A)$ (see main text). \textbf{E-H}: Species abundance plotted against niche values. (y-axis clipped at $N=15$ to emphasize abundance structure in the gaps between high-abundance species.) Clusters are quite distinctive in the noise-free scenario (E). In the noisy scenarios (F-H) the pattern is less regular but clustering still occurs---all examples shown are significant at statistical level $p\le 0.01$. Legend shows the estimated number of clusters and the z-score of the clustering statistic.} \label{FigS1_MultiDCoreTail}
\includegraphics[width=1\textwidth,angle=0]{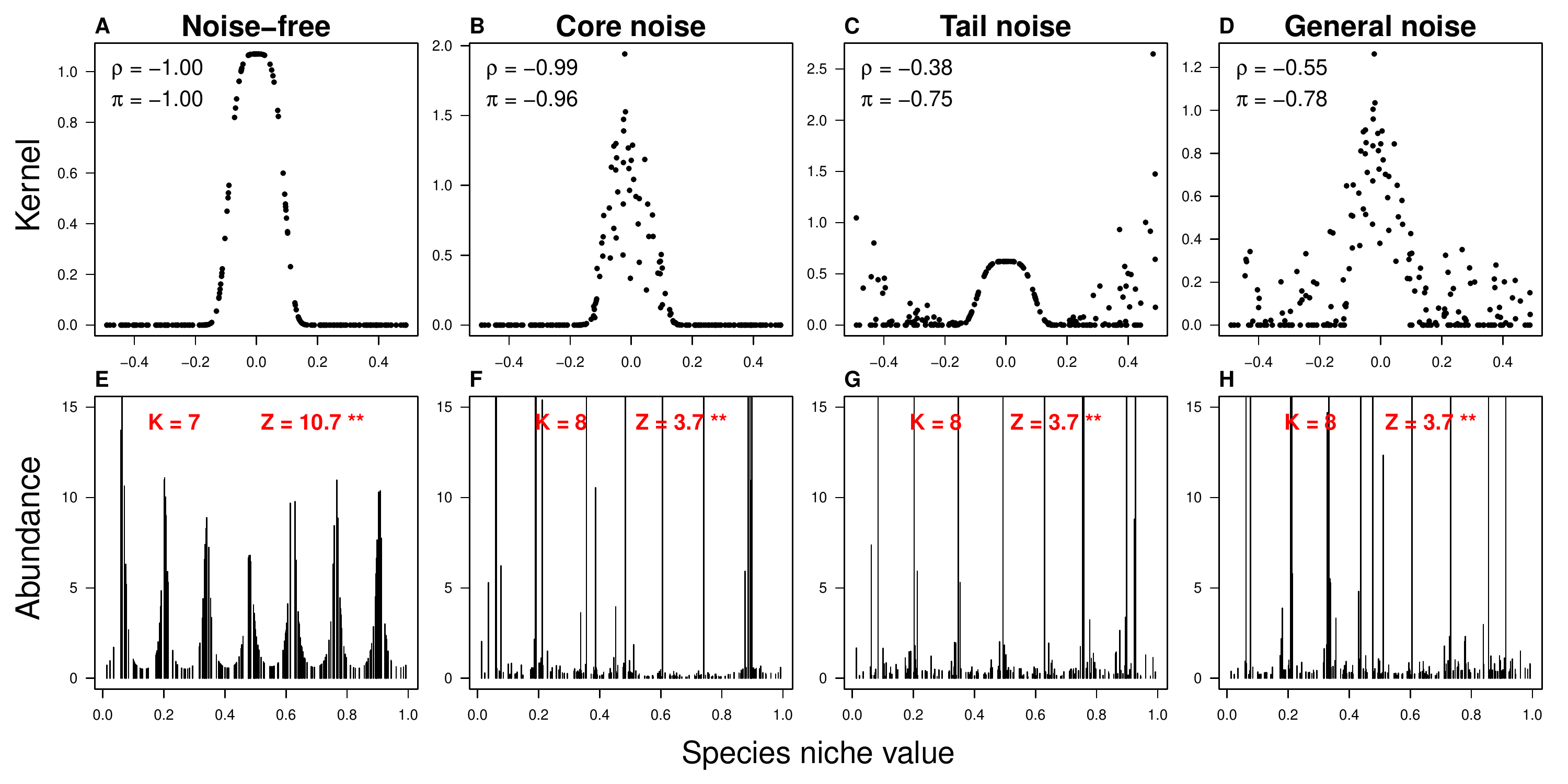}
\centering
\end{figure}

In the main text we showed how process noise arises from incomplete access to full niche space in an example where niche space is two-dimensional. Here we generalize this model, assuming an indeterminate number of unseen niche axes and other contributing factors which collectively add noise to the relationship between competition and x-niche differences. We thus set 
\[
A_{ij} = \textrm{exp}(-(|x_i-x_j|/w)^4) + \varepsilon_{ij},
\]
where the noise term $\varepsilon_{ij}$ is a normally distributed random variable with mean 0. We vary the magnitude or amplitude of the noise by tuning the variance parameter. By analogy with the two-dimensional scenario, where we observed that different types of relationship between the niche axes (complementary or essential) lead to more noise in specific regions of the kernel, here we consider three cases: noise can be stronger between species with small x-niche differences, large x-niche differences, or irrespective of x-niche differences. We call these core noise, tail noise, and uniform noise, in reference to the part of the kernel where most of the noise is concentrated. We achieve these different cases by allowing the variance of the noise to depend on distances $d_{ij}$ on the x-niche axis (see Fig. \ref{FigS1_MultiDCoreTail}B-D for visual representations of each). Because a noisy kernel is unique, we simulated 100 replicates (runs) for each noise type and magnitude to ensure robust results.

We impose nonnegative competition coefficients between all species pairs, and positive intraspecific competition for all species in order to ensure self-regulation and thus avoid unchecked growth. (If the noise term brings the coefficient below zero, we set it to zero if off-diagonal, or redraw it otherwise.) Once all competition coefficients are drawn we rescale the matrix by its mean, $A\rightarrow \mu_A\;A/\overline{A}$, in order to ensure that the average competitive impact between species pairs is the same across runs and scenarios. The rescaled  mean $\mu_A$ affects final community size. We use $\mu_A=0.2$, which combined with our other parameter choices leads to community sizes between 600 and 35,000 individuals, depending on the noise scenario and magnitude. We use a circular niche axis to avoid edge effects. This isolates the contribution of niche differences to pattern (further testing revealed this assumption is not critical for the validity of our results). 

Fig.~\ref{FigS1_MultiDCoreTail} shows examples of kernels and simulation outcomes in each case. When there is no noise, the kernel is perfectly monotonic on x-niche distances (Fig.~\ref{FigS1_MultiDCoreTail}A), and the resulting community is visibly clustered (Fig.~\ref{FigS1_MultiDCoreTail}E). As process noise is added in the form of core, tail, and uniform noise, the kernel lose monotonicity in the respective region while retaining a statistical trend (Fig.~\ref{FigS1_MultiDCoreTail}B-D). The resulting communities, despite the noise, are still clustered, albeit less distinctively so (Fig.~\ref{FigS1_MultiDCoreTail}F-H). Notice the parallels between core (tail) noise in this generalized process noise model and complementary (essential) niche axes in the two-dimensional model (cf. Fig.~\ref{FigS1_MultiDCoreTail} and Fig.~3 in the main text).

\begin{figure}[H]
\caption{\textbf{Probability of clustering by magnitude of noise}. The probability of clustering is estimated as the percentage of runs of each noise scenario where clustering was statistically significant ($p\le 0.05$), out of a total 100 replicates each. The noise is measured as the degree of randomness between the kernel and the x-niche axis (process noise) or the trait axis (measurement noise), using two indices: \textbf{A}: Monotonic decline of competition with species differences, measured with Spearman's rank correlation coefficient $\rho$. \textbf{B}: Core-tail structure, defined as the difference between the proportion of tail and core elements that exceed the kernel mean. Scenarios shown are process noise with complementary niche axes (blue curves), process noise with essential niche axes (red), and measurement noise (black). Compare with Fig.~3 in the main text.} \label{FigS2_DiagPlots}
\includegraphics[width=.75\textwidth,angle=0]{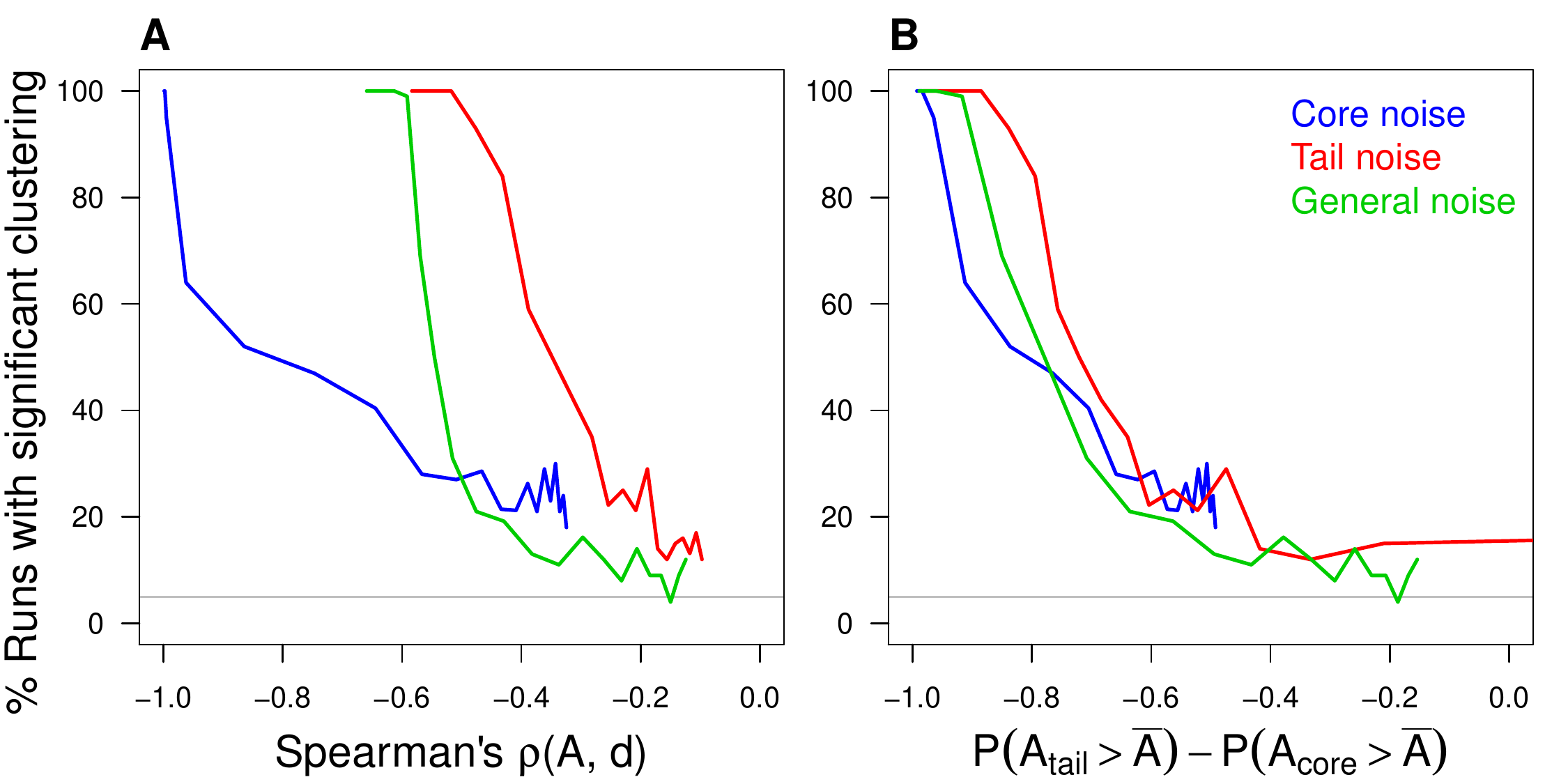}
\centering
\end{figure}

The probability of clustering decreases as the noise increases (Fig.~\ref{FigS2_DiagPlots}). For the same degree of overall monotonic relationship between competition and x-niche differences, core noise seems to degrade pattern the most, and tail noise the least, with uniform noise in between (cf. blue, red, and green curves in Fig.~\ref{FigS2_DiagPlots}A). Again, this is similar to what we observed in the two-dimensional case regarding complementary and essential niche axes. The core-tail structure predicts the probability of clustering consistently, whether the noise is in the core, tail, or across the kernel (Fig.~\ref{FigS2_DiagPlots}B). This is also similar to our observations for the two-dimensional model (cf. Fig.~\ref{FigS2_DiagPlots} and Fig.~4 in the main text).

The similarity between the results of our two-dimensional niche model and our generalized noise model indicates that higher niche dimensions add no qualitatively new phenomena to pattern along a single niche axis. The fact that uniform noise shows intermediate behavior between core and tail noise suggests that process noise originating from a combination of complementary and essential ``hidden'' niche axes will produce intermediate effects between strictly complementary and strictly essential niche axes.

\section{Estimating the niche value through multiple proxy traits} \label{appMultitrait}
In the main text we wrote 
\begin{equation}
t_i^a=b^a x_i+c^a+\varepsilon_i^a, \label{eq}
\end{equation}
where $t_i^a$ is species $i$'s trait value on trait axis $a$, $x_i$ is species $i$'s niche strategy, $b^a$ and $c^a$ are constants, and $\varepsilon_i^a$ is the measurement noise. Written this way, it looks like we are assuming the niche causes the trait. However, this is not so. 

In general, species $i$'s niche strategy along a niche axis will be a function of its trait values along $n$ functional trait axes: $x_i = f(T_i^1,T_i^2,\dots,T_i^n)$. Our example in the main text corresponds to the particular case where that function is linear in all of the trait values, $x_i=k^0+\sum_{j=1}^n k^j T_i^j$. This expression can be easily inverted to write the species trait along a given trait axis as a linear function of its niche strategy and all other trait values: $T_i^a=1/k^ax_i-k^0/k^a-\sum_{j\neq a}k^j/k^a T_i^j$. This makes it clear that the measurement error $\varepsilon_i^a$ between trait axis $a$ and the niche axis in Eq.~\ref{eq} stands for the combined contribution of the unmeasured trait axes. Assuming no correlation between trait values along different trait axes, we can find a linear transformation of $T_i^a$ that recovers Eq.~\ref{eq}, with $\underset{i}{\textrm{Mean}}(\varepsilon_i^a)=0$ and $\underset{i}{\textrm{Var}}(\varepsilon_i^a)=\sigma^2_a$. Parameter $\sigma_a$ reflects the amplitude of the noise between trait axis $a$ and the niche axis $x$. This noise amplitude will be small when the contribution of the unmeasured traits is small.

\begin{figure}[H]
\caption[]{Example showing relationship between the trait value of each species (x-axis) and the niche value (y-axis) for 8 proxy traits that predict the niche value with varying degrees of precision, plus the first principal component (ninth panel). The legend in each panel shows the coefficient of determination between the niche and trait values $R^2$, the estimated number of species clusters along that trait axis K, and the clustering z-score Z, with the number of stars indicating significance (2 stars for $p\le 0.01$, 1 star for $p\le 0.05$, zero stars for $p>0.05$).} \label{fig4}
\includegraphics[width=.8\textwidth,angle=0]{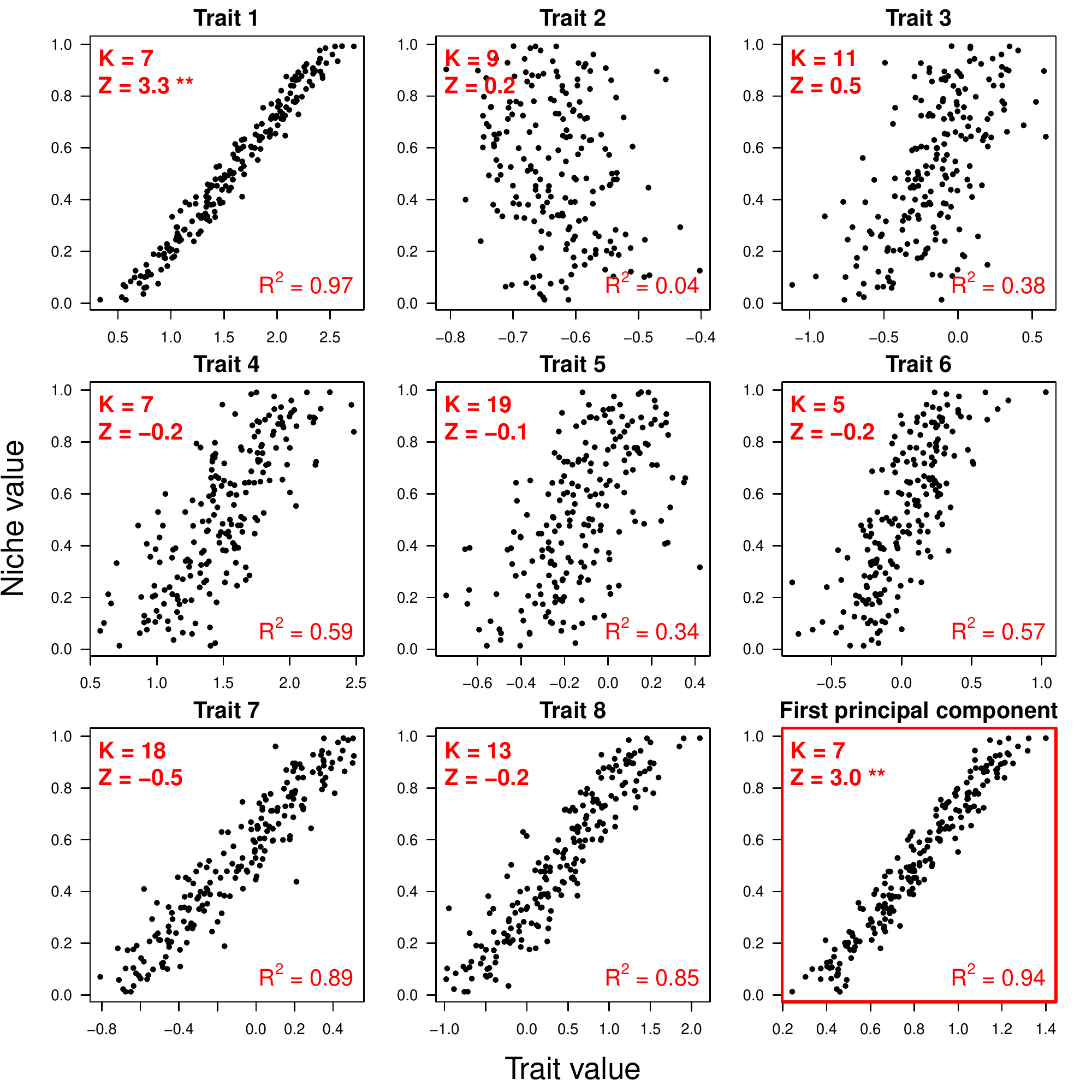}
\centering
\end{figure}

In the main text we assume furthermore that the noise is normally distributed, unbiased, and uncorrelated. In other words, $\varepsilon_i^a \sim N(0,\sigma_a)$, $\textrm{cov}(x_i,\varepsilon_i^a) = 0$, and $\textrm{cov}(\varepsilon_i^a,\varepsilon_j^b)=1$ if $i=j$ and $a=b$, and 0 otherwise. In these circumstances, the first principal component of such a set of proxy traits will typically predict species niche values better than most proxy traits individually. An example is shown in Fig.~\ref{fig4}.

Consider the special case where the expected value of each trait is the true niche value ($b^a=1$ and $c^a=0$, for each $a=1,2,\dots,L$). The coefficient of determination between the niche axis and trait axis $a$ is defined as $R^2_{x,a}=\frac{\textrm{cov}^2(x,\,t^a)}{\sigma_x^2\sigma_a^2}$, where $\sigma_x^2$, $\sigma_a^2$, and $\textrm{cov}(x,t^a)$ are respectively the variance in niche values, variance in trait values, and covariance between niche and trait values across species. Substituting we get 
\begin{eqnarray}
R^2_{x,a}&=&\frac{\textrm{cov}^2(x,x+\varepsilon^a)}{\sigma_x^2\sigma_a^2} \nonumber\\
&=&\frac{\textrm{cov}^2(x,x)}{\sigma_x^2(\sigma_x^2+\sigma_a^2)}\nonumber\\
&=&\frac{\sigma_x^2}{\sigma_x^2+\sigma_a^2}\nonumber
\end{eqnarray}
We define a species' consensus trait $z_i$ as the average across its proxy traits: $z_i=1/L \sum_a t_i^a=x_i + \sum_a \varepsilon_i^a/L=x_i+\overline{\varepsilon^a_i}$. It is also a random variable whose expected value is the true niche value, with the advantage that it has a lower-magnitude error: $\overline{\varepsilon^a_i}\sim N(0, \sqrt{\overline{\sigma_a^2}/L})$. By analogy with the calculation above, the $R^2$ between the niche values and the consensus trait is $R^2_{x,z}=\frac{\sigma_x^2}{\sigma_x^2+\overline{\sigma_a^2}/L}$. Therefore the consensus trait will have a higher $R^2$ against the niche axis than a given trait $t^a$ if $\overline{\sigma^2_a}/L<\sigma^2_a$. If all traits have the same noise amplitude, then $\overline{\sigma^2_a}=\sigma^2_a$ and the condition is trivially met for any number of trait axes $L>1$. When the noise amplitudes differ across traits, the consensus trait will still have a higher coefficient of determination that then ``typical'' trait with noise amplitude $\overline{\sigma^2_a}$.

In the more general case where $b^a$ and $c^a$ differ across proxy traits, the first principal component replaces the simple trait average as the best predictor of the niche value. In Fig.~\ref{fig4} we show results for an example with 8 proxy traits where the constants $b^a$ and $c^a$ were respectively drawn from a normal distribution centered at 1 and 0: $b^a\sim N(1,\sigma_b)$, $c^a\sim N(0,\sigma_c)$. We used $\sigma_b=\sigma_c=0.5$. The amplitude of the measurement noise $\sigma_a$ for each trait was drawn from a uniform distribution between 0.05 and 0.25. 

In the example shown in Fig.~\ref{fig4}, most trait axes fail to reveal the existing clustering pattern along the niche axis, even those with high $R^2$ (Trait 7 and Trait 8). In contrast, the first principal component is clustered (clustering z-score $= 3.0$, p-value $< 0.01$). Note that in this example the first principal component is a better predictor of the niche than seven of the eight traits ($R^2=0.94$), but is a poorer predictor than Trait 1, which is tightly linked to the niche ($R^2=0.99$) and is the only clustered trait (z-score $= 3.3$, p-value $< 0.01$). In general, the first principal component will typically have a higher coefficient of determination with the niche axis than individual trait axes, and is also more likely to be clustered.

\section{Clustering metric --- the gap statistic via k-means clustering}\label{appMetric}
Here we describe a metric to identify and quantify clustering between species on a niche or trait axis\footnote{The method also applies in general to a niche or trait space with any number of dimensions, as long as that space allows a meaningful definition of ``distance'' between species---such as Euclidean distance.}, based on a method developed by \citet{Tibshirani2001}. The metric receives as input the species abundances and their niche (trait) values, and then estimates the number of clusters that best fits the data. It also provides an index---the gap statistic---that quantifies the degree to which the species assemblage fits that number of clusters better than a set of reference (null) assemblages. 

For each candidate number of clusters $k$ within a provided range $[k_\textrm{min}, k_\textrm{max}]$, we calculate the goodness of fit of $k$ clusters to our species assemblage, $f_k$, as well as the goodness of fit of $k$ clusters to each of the null assemblages $\tilde{f}_{k,i}$, where $i=1,2,\dots,N_\textrm{nulls}$ and $N_\textrm{nulls}$ is the total number of null assemblages. We then record the difference (the gap) between the former and the average of the latter: $\Delta_k=f_k-\overline{\tilde{f}_k}$. The estimated number of clusters $\hat{k}$ is the $k$ that maximizes the gap, $\Delta_{\hat{k}}=\underset{k}{\textrm{max}}(\Delta_k)$, and the maximum gap itself is the gap statistic, $Gap=\Delta_{\hat{k}}$. 

The method is quite general, and both the null assemblages and the measure of goodness of fit must be specified. We generate our null assemblages by shuffling abundances among species, i.e. we randomly redistribute observed species abundances to observed niche (trait) values. This keeps both the set of abundances and the set of traits, while randomizing the map between them. 

As for the goodness of fit, following \citep{Tibshirani2001} we use the k-means clustering algorithm (see \citealt{MacQueen1967}). This method partitions individuals into a specified number of groups such that the so-called dispersion between individuals within the same group is minimized. Let $D_r$ be the total pairwise squared distance between members of group $r$, that is, $D_r=\sum_{i,j}(x_i-x_j)^2$, where $x_i$ is the niche or trait value of individual $i$. The dispersion $W_k$ is then the sum of those numbers across all $k$ groups, $W_k=\sum_{r=1}^k D_r$. The k-means algorithm consists in assigning each individual to one of the $k$ groups so that $W_k$ is as low as possible\footnote{Dispersion can be defined in different ways, depending on how one wishes to weigh the number of members per group. While \citet{Tibshirani2001} choose to scale the squared distances by the number of individuals in the group, \citet{Yan2007} argue that scaling it by the number of pairs in the group is a a better choice in some cases. In our study we saw no qualitative difference between those different scaling options, and opted to use the simpler unweighted sum of squared distances.}$^,$\footnote{In our model, all conspecific individuals have the same niche (trait) value, and therefore necessarily belong in the same cluster. Species are therefore trivial clusters of individuals. Ecologists are of course interested in clustering beyond the obvious grouping of individuals into species.}. 

We then set $f_k=\textrm{log}(1/W_k)$ as the goodness of fit, and feed it into the gap statistic algorithm to estimate the number of clusters $k$. Notice that it does not make sense to compare the goodness of fit $f_k$ directly across different $k$’s because $f_k$ necessarily increases with $k$, as the average within-cluster distance is always lower for higher numbers of clusters. By comparing against null assemblages, the gap method finds the biggest increase in goodness of fit \emph{beyond} what is expected from the increase in $k$. The reason we use $\textrm{log}(1/W_k)$ rather than simply $1/W_k$ is that the expected increments in $1/W_k$ with increasing $k$ are multiplicative rather than additive. See \citep{Tibshirani2001} for more details. Our metric is then defined as $Gap=\underset{k}{\textrm{max}}\left(\textrm{log}(1/W_k) - \overline{\textrm{log}(1/\tilde{W}_k)}\right)$, and the value of $k$ at which the quantity in the parenthesis is maximal is the estimated number of clusters $\hat
{k}$. 

If we knew the expected distribution of this statistic under the null hypothesis where species niches (traits) and abundances are unrelated to each other, we could stop here, as the value of $Gap$ would tell us about both degree of clustering and statistical significance. Because we do not know that distribtution, we must compare it with a null set of values. We therefore calculate the metric on each of our null assemblages, and from that we obtain a z-score and a p-value. Those are defined as $z=(Gap - \overline{\tilde{Gap}}) \,/ \, \sigma(\tilde{Gap})$ and $p=\frac{1}{N_\textrm{nulls}}\sum_{i=1} ^{N_\textrm{nulls}} I(\tilde{Gap}_i\ge Gap)$ where $\tilde{Gap}_i$ is the gap statistic obtained for the $i$-th null assemblage, $\overline{\tilde{Gap}}$ and $\sigma(\tilde{Gap})$ are the mean and standard deviation across those values, and $I$ is the index function, equal to 1 if its argument is true and zero otherwise. The z-score and p-value quantify respectively the degree to which the gap statistic returned by our species assemblage exceeds null expectations, and whether that result is statistically significant.

\section{Kernel structure in the trait-noise scenario} \label{appStructure}
\begin{figure}[H]
\caption[]{Comparison between the competition kernel from the core-noise scenario and the trait-noise scenario. \textbf{Left}: Core-noise competition coefficients between species pairs plotted against their respective niche values, with darker shades of gray representing stronger competition. Coefficients tend to be higher when species have similar niche values (the top-left and bottom-right corners are continuations of the diagonal, as the niche axis is circular). The noise is manifested in the fine-grained pixelation of the grey band, and lacks any visible structure. \textbf{Right}: Trait-noise competition coefficients plotted against trait values. As in the core-noise scenario, the noise is concentrated in the diagonal band where competition is stronger. However, unlike the core-noise scenario, the matrix is symmetric and the pixel sizes are larger, indicating autocorrelation in the noise.} \label{supfig_corevtrait}
\includegraphics[width=1\textwidth,angle=0]{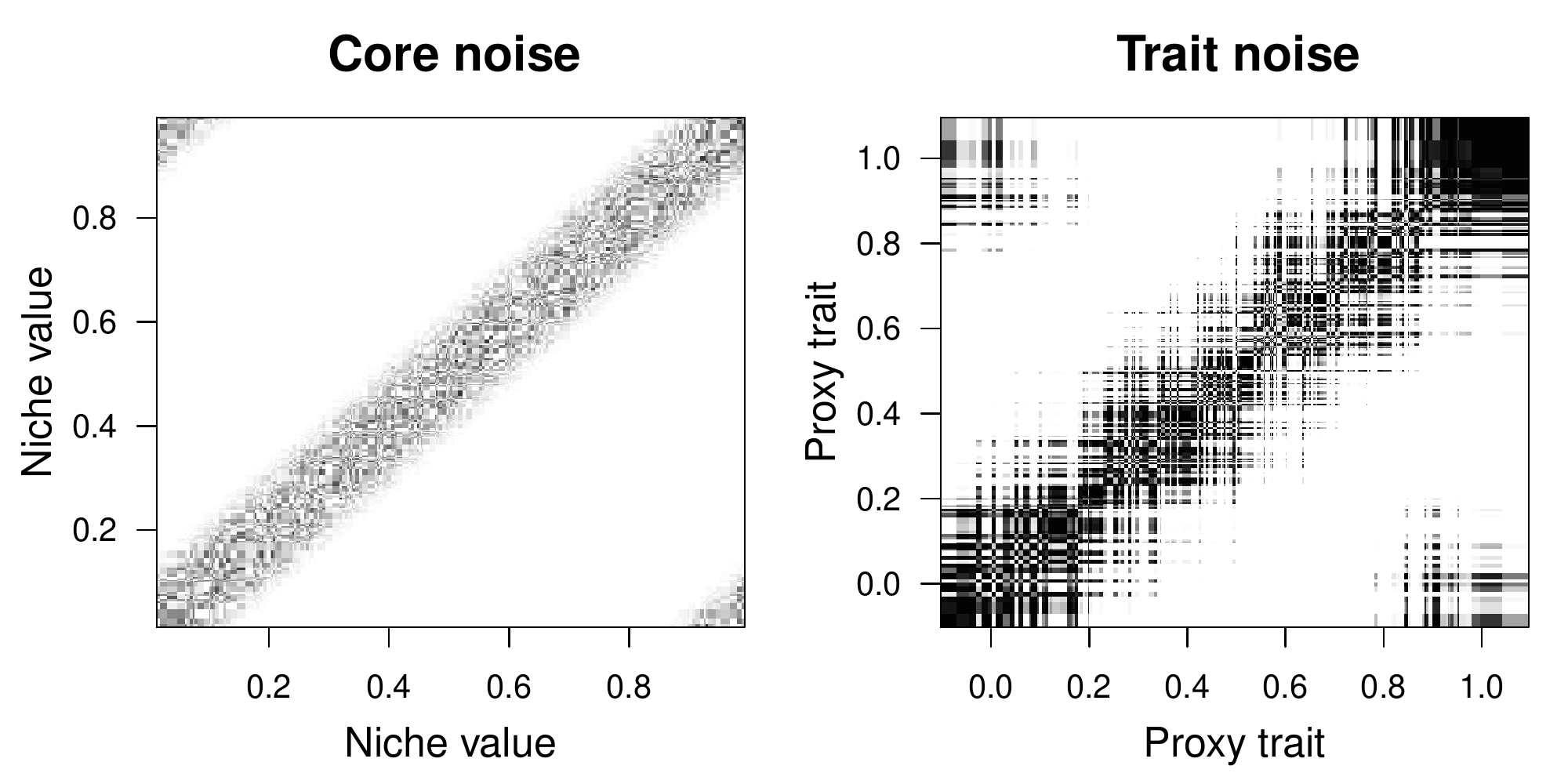}
\centering
\end{figure}

In the trait-noise scenario, the competition coefficients plotted against proxy trait values resemble the core-noise kernel, with most of the noise occurring for pairs with small trait difference (compare Figs.~3A and 5H in the main text). However, a closer look reveals that the measurement-noise kernel is more structured (Fig.~\ref{supfig_corevtrait}): unlike the core-noise scenario, the matrix is symmetric and the noise is autocorrelated. We note that there is a permutation of the rows and columns of the competition matrix in the trait noise scenario that restores the noise-free matrix. This is not the case with any of the kernel noise scenarios presented in the main text.

